\documentclass[11pt,a4paper]{article}
\usepackage[T1]{fontenc}
\usepackage[utf8]{inputenc}
\usepackage{lmodern}
\usepackage[margin=2.5cm]{geometry}
\usepackage{amsmath,amssymb,mathtools}
\usepackage{graphicx}
\usepackage{xcolor}
\usepackage{booktabs}
\usepackage{makecell}
\usepackage{enumitem}
\usepackage[font=small,labelfont=bf]{caption}
\usepackage[numbers,sort&compress]{natbib}
\usepackage{authblk}
\usepackage[colorlinks=true,linkcolor=blue!50!black,citecolor=blue!50!black,urlcolor=blue!50!black]{hyperref}

\newcommand{\fA}{f_{A_0}}
\newcommand{\fB}{f_{B_0}}
\newcommand{\chiA}{\chi_A}
\newcommand{\chiB}{\chi_B}
\newcommand{\Qt}{\widetilde{Q}}

\title{Explaining intersectional segregation\\in multi-attribute social networks}
\author[1]{Clara Eminente}
\author[1,*]{Elisa Omodei}
\author[2,*]{Michele Starnini}
\affil[1]{Department of Network and Data Science, Central European University, 1100 Vienna, Austria}
\affil[2]{Department of Engineering, Universitat Pompeu Fabra, 08018 Barcelona, Spain}
\affil[*]{Corresponding authors: michele.starnini@upf.edu; omodeie@ceu.edu}
\date{}

\begin{document}
\maketitle

\begin{abstract}
Segregation in students' friendships shapes their access to information and opportunities, and schools can reduce it by setting the composition of their classes. Yet segregation is usually measured one attribute at a time, although attributes such as migration background and sex are often correlated, so a preference over one can create segregation on the other. A single-attribute reading then attributes this induced segregation to the wrong preference and cannot identify which part of it a school can remove. In a network model with preferences over two correlated attributes, we derive in closed form the Coleman index, a standard segregation measure, separating the segregation due to each preference from the segregation it induces on the other attribute. In friendship networks of school classes in four European countries, ignoring sex inflates the estimated preference over migration background, increasingly so as the two attributes are more correlated. The closed form also shows what class composition can achieve. When students have no preference over one attribute, the segregation induced on it can be removed, whereas segregation set by a homophilic preference over the other cannot be changed. When both preferences are active, splitting a school with equal-sized groups and independent attributes reduces segregation only if the preferences pull in opposite directions. Judging a split by the school-wide average can hide classes that become more segregated. In the data, the split derived from the closed form removes induced segregation, and splitting does not lower segregation where both preferences are homophilic, as predicted.
\end{abstract}

\noindent\textbf{Keywords:} segregation; homophily; intersectionality; school friendship networks; classroom composition

\section{Introduction}
Segregation in social networks, the tendency to form ties mostly within one's own group, shapes educational and economic outcomes. In-school segregation reduces the educational attainment and long-run earnings of minority students \citep{mickelson2001subverting, card2007racial, chetty2016effects}, and economic connectedness, measured on hundreds of millions of Facebook friendships, is among the strongest predictors of upward mobility \citep{chetty2022social}. Segregated networks concentrate information, referrals and social influence within advantaged groups \citep{jackson2025inequality, dimaggio2012network}, affect the wellbeing of students \citep{boda2023ethnic}, and shape intergroup attitudes, as children with few cross-race friendships are more likely to attribute negative intent to minority peers \citep{killen2022reducing}.

Understanding what produces segregation requires looking at more than one attribute at a time. Individuals carry several characteristics, and the correlation between two of them in a population, their consolidation, shapes friendship segregation \citep{blau1977inequality, blau2018crosscutting}. Across 42 attribute combinations in 524 school classes, gender consolidation is the most powerful driver of friendship segregation \citep{boller2025friendship}. Ethnicity ``borrows groupness'' from gender when the two attributes are aligned \citep{kroneberg2021ethnicity}, an effect that requires strong homophily on the second attribute \citep{kroneberg2024structural}. Interethnic friendships are less common in classrooms where ethnic groups differ in socioeconomic status \citep{zhao2023superdiversity}, and racial and socioeconomic preferences acting jointly drive about 37\% of socioeconomic segregation in US high schools \citep{rosche2024micro}, although homophily on different dimensions can also act as a substitute rather than a complement \citep{block2014multidimensional}. Despite the extensive empirical evidence of its effects, an analytical account of how consolidation produces segregation is still missing.

\begin{figure}[t]
\centering
\includegraphics[width=0.6\textwidth]{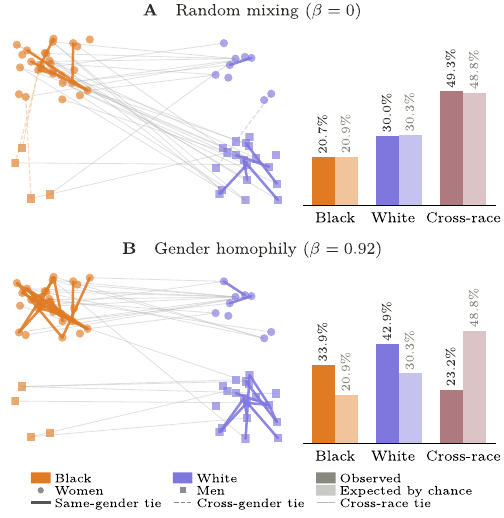}
\caption{\textbf{Preferences over gender create segregation by race.} A simulated school of $N=60$ students in two racial groups with different gender compositions: 92\% of the Black students and 16\% of the White students are women ($\fA=0.45$, $\fB=0.5$, $k=0.92$). Nobody has a racial preference ($\alpha=0$). Bars: shares of within-Black, within-White and cross-race ties, observed (dark) and expected by chance, mirroring the population shares (light). (A) Without a gender preference ($\beta=0$), the observed shares match the expected ones. (B) With gender homophily ($\beta=0.92$), cross-race ties fall to 23\% of all ties, against 49\% expected by chance. }%
\label{fig:induced}
\end{figure}

In this paper we show how consolidation produces segregation, using a standard measure of network segregation, the Coleman index \citep{coleman1958relational}. The index compares the fraction of ties within groups with the fraction expected by chance: it is zero when ties form by chance, positive under homophily and negative under heterophily, and it is computed one attribute at a time. We calculate it in a multi-attribute model (MAM), in which the probability that two students become friends is the product of two terms, one for each attribute (race and gender in our running example), each set by the preference over that attribute \citep{martin2026simple}. Our main result is the Coleman index of each attribute of this model in closed form. 
Through this closed form we show that segregation measured on race has two sources: the preference over race, and the preference over gender acting through the correlation between race and gender, which we call induced segregation. Fig.~\ref{fig:induced} illustrates it: students have no racial preference, yet gender homophily reduces cross-race friendships, because the two racial groups have different gender compositions. A single-attribute reading of the index therefore attributes segregation to the wrong cause, reporting as a racial preference what is induced by the preference over gender.

The closed form also shows what a school can achieve by assigning students to classes, a lever to reduce segregation that Moody \citep{moody2001race} argues is more realistic than changing preferences, and that earlier work on attribute consolidation has studied only by simulation \citep{boller2025friendship}. When students have no racial preference and prefer friends of their own gender, classes in which race and gender are uncorrelated have no racial segregation, whereas gender segregation equals the gender preference in every class, so no assignment can reduce it. When both preferences matter, in a school 
with balanced, independent attributes, splitting reduces segregation only if the two preferences point in opposite directions. Away from independence, a split can lower the school-wide average of segregation while every class becomes more segregated on one attribute, a cancellation trap that, as we show, can be avoided using a class-by-class criterion.

We test these predictions on the friendship networks of 139 school classes from the Children of Immigrants Longitudinal Survey in Four European Countries (CILS4EU) \citep{kalter2016cils4eu}, collected in England, Germany, the Netherlands and Sweden. The CILS4EU data are widely used to study adolescent friendship networks, including ethnic homophily and its dependence on classroom composition \citep{smith2016ethnic}, ethnic boundary making \citep{kruse2019more} and gender segregation \citep{vit2026rigid}. Each student has two binary attributes, country of origin and sex; country of origin indicates migration background, that is, whether the student or a parent was born abroad. We use country of origin in place of race for two reasons. Race is among the attributes most studied in the homophily literature \citep{mcpherson2001birds}, but in Europe ethnic boundaries are drawn mainly along migration background and national origin. Racial categories are also rarely recorded, in part because European data-protection law treats racial or ethnic origin as a special category of personal data \citep{simon2012collecting, eu2016gdpr}, and European research on ethnic homophily, accordingly, relies on origin-based categories \citep{smith2014ethnic, leszczensky2019drives, dollmann2016examining}. Three findings emerge. First, in every class a model with both country of origin and sex describes who befriends whom better than a model with origin alone, and the latter assigns to origin the segregation that the preference over sex induces on it, the more so the more strongly origin and sex are consolidated. Second, in classes with almost no preference over one attribute, the equal-size split that makes the two attributes independent within each class reaches the predicted floor of segregation. Third, in the classes that are nearly balanced on both attributes and whose attributes are nearly independent, both preferences are homophilic and, as predicted, splitting does not reduce segregation.

\section{Results}

\subsection{A multi-attribute model}
To read the Coleman index of one attribute, we need a tie formation model in which the probability of a tie depends on both attributes of the two individuals. This subsection defines such a model: a population described by two correlated attributes, and a tie probability that combines one preference for each attribute.%

We consider a population of $N$ individuals, each of whom carries two binary attributes: attribute~$A$, with groups $A_0$ and $A_1$, and attribute~$B$, with groups $B_0$ and $B_1$. In our running example, $A$ is race and $B$ is gender. We write $\fA$ for the fraction of individuals in group~$A_0$ and $\fB$ for the fraction in group~$B_0$, and we label the groups so that $A_0$ and $B_0$ are the smaller ones ($\fA,\fB\leq 1/2$). The two attributes can be correlated. We describe their correlation by the consolidation~$k$, the fraction of the members of~$A_0$ who belong to~$B_0$; both conventions are those of Martin-Gutierrez et al.\ \citep{martin2025intersectional}. When $k=\fB$, the share of~$B_0$ among the members of~$A_0$ equals its share in the whole population, and the two attributes are independent; the further $k$ is from~$\fB$, in either direction, the more strongly the two attributes are consolidated; in particular they are correlated if $k>\fB$ and anticorrelated if $k<\fB$. %
Together, $\fA$, $\fB$ and~$k$ determine the fraction of individuals of each of the four types $(A_a,B_b)$ (Materials and Methods).

Ties form independently between pairs of individuals. The probability of a tie between individuals $i$ and~$j$ is proportional to the product of two terms, one for each attribute:
\begin{equation}\label{eq:link_prob}
p(i,j) \propto H^A_{A(i),A(j)}\, H^B_{B(i),B(j)}, \quad
H^X = \tfrac{1}{2}\left(\begin{smallmatrix} 1+x & 1-x \\ 1-x & 1+x \end{smallmatrix}\right),
\end{equation}
where $A(i)$ and $B(i)$ are the groups of~$i$, and $x=\alpha$ for attribute~$A$ and $x=\beta$ for attribute~$B$. The term for attribute~$A$ is thus $(1+\alpha)/2$ if the two individuals belong to the same $A$-group and $(1-\alpha)/2$ otherwise, and similarly for~$B$. The preferences $\alpha$ and~$\beta$ lie between $-1$ and~$1$. A positive~$\alpha$ means homophily over~$A$: ties within an $A$-group are more likely than ties across the two groups. A negative~$\alpha$ means heterophily: ties across groups are more likely. When $\alpha=0$, attribute~$A$ plays no role in the formation of ties. The same holds for~$\beta$ and attribute~$B$. We call the model of Eq.~\ref{eq:link_prob} the multi-attribute model (MAM). It is a version of the model of Martin-Gutierrez et al.\ \citep{martin2026simple}, who compared several models of how multidimensional identities shape social networks and found that this multiplicative model reproduces empirical networks best. Our version has a single preference parameter for each attribute instead of a full matrix of preferences; this restriction is what makes the model analytically tractable (Materials and Methods).

The product form means that a meeting between two individuals is judged once on each attribute, and that the two judgments are independent. In this reduced form, $\alpha$ and~$\beta$ summarize, for each attribute, the net effect of everything that makes ties within its groups more or less likely than chance, preferences included.

\subsection{The Coleman index of one attribute mixes two preferences}
Segregation on an attribute is usually summarized by a single number, computed from the ties within and between the groups defined by that attribute. We ask whether this number, the Coleman index, reveals the preference over the attribute when individuals also have a preference over a second, correlated attribute. We calculate the index in closed form in the MAM and show that it depends on both preferences.

We use the original definition of the index \citep{coleman1958relational, bojanowski2014measuring} and state it for attribute~$A$; a group is then the set of individuals sharing the same $A$-group, whatever their $B$-group. Let $w_A$ be the fraction of ties that fall within a group, and $w_A^*$ the fraction expected if every individual kept its number of ties but chose its partners at random. We call $w_A^*$ the Coleman baseline (formula in Materials and Methods). The Coleman index of attribute~$A$ is
\begin{equation}\label{eq:CI}
\chiA = \begin{cases}
\dfrac{w_A - w_A^*}{1 - w_A^*}, & w_A \geq w_A^*,\\[6pt]
\dfrac{w_A - w_A^*}{w_A^*}, & w_A < w_A^*.
\end{cases}
\end{equation}
The first line applies when within-group ties are at least as frequent as the baseline predicts (the homophilic regime), the second when they are less frequent (the heterophilic regime). The index is zero when the fraction of within-group ties equals the baseline, one under complete segregation ($w_A=1$), and minus one when no tie falls within a group ($w_A=0$). The index can be measured on a network, by counting its ties, or calculated in a network model, by replacing the counts with their expected values; only in the second case is its dependence on both preferences made explicit. %

In the MAM, the index is calculated from the expected densities of ties within and between groups (Materials and Methods), and its closed form depends on how attribute~$B$ is distributed within each $A$-group. We describe this distribution by the $B$-imbalance $\delta_a$ of group~$A_a$: the fraction of the population that belongs to $A_a$ and to~$B_0$, minus the fraction that belongs to $A_a$ and to~$B_1$. It is zero when $A_a$ has as many members in~$B_0$ as in~$B_1$, positive when $B_0$ is over-represented in~$A_a$, and negative when $B_1$ is; Materials and Methods gives it in terms of $\fA$, $\fB$ and~$k$. In the school of Fig.~\ref{fig:induced}A--B, where 92\% of the Black students and 16\% of the White students are women, the two imbalances are $\delta_0=0.38$ and $\delta_1=-0.38$. In the homophilic regime, the index reads
\begin{equation}\label{eq:chiA_closedform}
\chiA = \frac{\alpha\,\fA(1-\fA) + \beta\,Q_A}{\fA(1-\fA) + \beta\,\widetilde{Q}_A},
\end{equation}
where
\begin{equation}\label{eq:QA}
\begin{aligned}
Q_A &= \tfrac{1+\alpha}{2}\bigl[(1-\fA)\,\delta_0^2+\fA\,\delta_1^2\bigr]-\tfrac{1-\alpha}{2}\,\delta_0\,\delta_1,\\
\widetilde{Q}_A &= \tfrac{1+\alpha}{2}\bigl[(1-\fA)\,\delta_0^2+\fA\,\delta_1^2\bigr]+\tfrac{1-\alpha}{2}\,\delta_0\,\delta_1
\end{aligned}
\end{equation}
are quadratic forms in the two $B$-imbalances (SI Appendix, section~\ref{S-si:closedform}).

The index measured on networks simulated from the MAM agrees with Eq.~\ref{eq:chiA_closedform} over the whole range of consolidation (SI Appendix, Fig.~\ref{S-fig:validation}; Materials and Methods). 

The numerator of Eq.~\ref{eq:chiA_closedform} is proportional to the expected excess of within-group ties over the baseline, and it is the sum of two terms. The first, proportional to~$\alpha$, is the direct effect of the preference over~$A$. The second, proportional to~$\beta$, is the indirect effect of the preference over~$B$: it is where that preference enters the index of~$A$, and it depends on the composition of the population through the $B$-imbalances. With no preference over~$B$ ($\beta=0$), only the direct term is left, the denominator reduces to $\fA(1-\fA)$, and the index equals~$\alpha$. The preference over~$B$ also enters the denominator, through~$\widetilde{Q}_A$, so the indirect term alone does not give the effect of that preference. When the two attributes are independent, for instance, the indirect term is in general not zero, but it is offset by the term $\beta\widetilde{Q}_A$ of the denominator, and the index equals~$\alpha$ (SI Appendix, section~\ref{S-si:chiA_minus_alpha}). The effect of the preference over~$B$ is given by the difference between the index and~$\alpha$, Eq.~\ref{eq:chiA_minus_alpha}. In the heterophilic regime, the index has the same numerator over a different positive denominator, and it can be expressed as Eq.~\ref{eq:chiA_closedform} multiplied by a positive factor (Materials and Methods). The index of~$B$ follows from Eqs.~\ref{eq:chiA_closedform} and~\ref{eq:QA} by exchanging the roles of the two attributes (Materials and Methods).

\subsection{Correlated attributes can induce segregation}
The index of~$A$ can be read as the preference over~$A$ only in special cases. 
This subsection characterizes these cases and shows that, in all others, the difference is equal to the segregation induced by the preference over~$B$.

How the two preferences combine is clearest from the difference between the index and the preference over~$A$. In the homophilic regime, Eqs.~\ref{eq:chiA_closedform} and~\ref{eq:QA} give (SI Appendix, section~\ref{S-si:chiA_minus_alpha})
\begin{equation}\label{eq:chiA_minus_alpha}
\chiA-\alpha \;=\; \frac{\beta\,(1-\alpha^2)\,\fA\,(k-\fB)\,\Delta}{\fA(1-\fA)+\beta\,\widetilde{Q}_A}.
\end{equation}
Here $\Delta=\delta_0-\delta_1$, the difference between the $B$-imbalances of the two $A$-groups, quantifies how much the two groups differ in their $B$-composition, and the denominator, the same as in Eq.~\ref{eq:chiA_closedform}, is positive. Since $1-\alpha^2$ and the denominator never vanish, the index differs from~$\alpha$ unless one of the three factors $\beta$, $k-\fB$ and~$\Delta$ vanishes. With homophily over~$A$ ($\alpha>0$), the index therefore equals the preference in three cases only.

The first case is when there is no preference over~$B$ ($\beta=0$): the index then equals~$\alpha$, whatever the composition of the population. This holds although groups rarely have the same number of ties, a phenomenon known as degree inequality. Under homophily, members of a larger group find more partners of their own type and form more ties, so that minorities tend to be less connected than majorities \citep{karimi2018homophily, currarini2009economic}; when attributes are correlated, preferences over one attribute also create degree inequality between the groups of the other \citep{martin2025intersectional}, and both mechanisms operate in the MAM (SI Appendix, section~\ref{S-si:gamma}). The Coleman baseline accounts for this: a group that sends a larger share of all ties is expected to have more within-group ties (Materials and Methods), and only the ties in excess of this expectation count as segregation. An index computed with the random-mixing value $\fA^2+(1-\fA)^2$, the fraction of within-group ties expected when all individuals have the same number of ties, would instead read these extra ties as segregation, and would exceed~$\alpha$ whenever the two groups differ in size (for instance, $0.67$ instead of $0.60$ for $\fA=0.2$ and $\alpha=0.6$; SI Appendix, section~\ref{S-si:randommixing}). The second case is when the two attributes are independent ($k=\fB$). The third is when $\Delta=0$, that is, when the two $A$-groups have the same $B$-imbalance, so that each accounts for exactly half of the $B$-imbalance of the population, $\delta_0+\delta_1=2\fB-1$.

Outside these three cases, the index also contains a contribution created by the preference over~$B$ and differs from~$\alpha$. We call this contribution induced segregation, and its size in the homophilic regime is given by Eq.~\ref{eq:chiA_minus_alpha}. Two ingredients are needed: individuals must have a preference over~$B$, and the two attributes must be correlated, so that the two $A$-groups differ in their $B$-composition. %
 Fig.~\ref{fig:induced}A--B illustrates the mechanism. When one racial group is mostly female and the other mostly male, gender homophily directs ties within racial groups, even though no one has a racial preference: in the simulated school of Fig.~\ref{fig:induced}B, cross-race ties fall to 23\% of all ties, against 49\% expected by chance. A single-attribute reading of the Coleman index attributes all the segregation on~$A$ to the preference over~$A$, and thus mistakes induced segregation for a preference.

Induced segregation is not a by-product of differences in the number of ties: such differences enter the index only through the Coleman baseline, which takes them into account, and, without a preference over~$A$, they require the two $B$-groups to differ in size. With no preference over~$A$ and two $B$-groups of the same size, every individual has the same expected number of ties, yet the index of~$A$ is positive with homophily over~$B$ whenever the attributes are correlated, as in Fig.~\ref{fig:induced}B (SI Appendix, section~\ref{S-si:equalties}). In general, when there is no preference over~$A$ ($\alpha=0$), the direct term vanishes and all the segregation on~$A$ is induced.

\subsection{A single-attribute model misattributes segregation in empirical school networks}
The closed form predicts that a single-attribute model misassigns to its only attribute the segregation induced by the other, omitted one.
We test this prediction on school friendship networks, by estimating the preferences of each class with and without the second attribute and by comparing the difference between the two estimates with the induced segregation that the closed form predicts.

We use the friendship networks of the Children of Immigrants Longitudinal Survey in Four European Countries (CILS4EU) \citep{kalter2016cils4eu}, a study of the integration of children of immigrants in England, Germany, the Netherlands and Sweden that started in 2010. Each network is a school class, and each tie is a friendship nomination, directed from the student who names a classmate to the classmate named. After filtering, the sample contains 139 classes (Materials and Methods). Every student has two binary attributes: country of origin (migration background) and sex. In each class, we label as~$A$ the attribute with the smaller minority group, as in the model, so that $A$ is country of origin in some classes and sex in others. These data also suit the study of class composition, because in European school systems students are assigned to classes by an administrator.

We estimate the preference parameters of the model fitted on each class by maximum likelihood, both in the MAM and in its single-attribute reduction (SAM) with country of origin as the attribute with active preference (Materials and Methods). A likelihood-ratio test favors the MAM over the SAM in every one of the 139 classes, at the 5\% significance level: in every class, a model with both attributes describes who befriends whom better than a model with origin alone. A model with both attributes is also better than a model with sex alone in 74 classes, about half of the sample (SI Appendix, section~\ref{S-si:replicas}). The MAM is a constrained version of the model of Martin-Gutierrez et al.\ \citep{martin2026simple}, whose unconstrained version (which we refer to as unconstrained MAM, uMAM) has a total of 8 preference parameters (4 per attribute), of which 7 are free (Materials and Methods). In 69 classes, about half of the sample, the uMAM does not fit the data significantly better than the MAM. Since the remaining analyses rely on the estimated preferences, we carry them out on these 69 classes, in which the constraints of the MAM do not significantly degrade the fit.

\begin{figure}[t]
\centering
\includegraphics[width=0.6\textwidth]{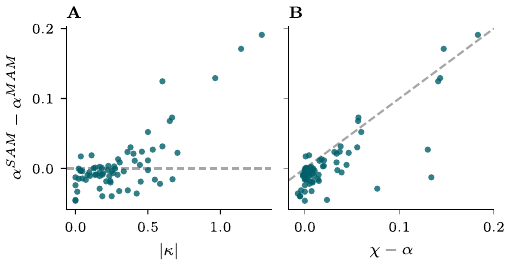}
\caption{\textbf{A single-attribute model misattributes segregation.} (A) Difference between the preferences over country of origin estimated with the single-attribute model (SAM) and with the multi-attribute model (MAM), $\alpha^{\rm SAM}-\alpha^{\rm MAM}$, against the absolute normalized consolidation $|\kappa|=|k-\fB|/\fB$, for the 69 classes in which the MAM fits as well as the unconstrained model; dashed line: zero. The difference grows with consolidation. (B) The same difference against the induced segregation $\chi-\alpha$ calculated from the closed form (Eq.~\ref{eq:chiA_minus_alpha} in the homophilic regime), with the composition of each class and the preferences estimated with the MAM; dashed line: the diagonal. Here $\alpha$ denotes the preference over country of origin, although origin is attribute~$A$ in some classes and attribute~$B$ in others; $|\kappa|=|F_{A_0,B_0}-\fA\fB|/(\fA\fB)$ is symmetric in the two attributes, so it does not depend on this labeling.}
\label{fig:data}
\end{figure}

If a single-attribute reading misattributes induced segregation, the SAM should estimate a stronger preference over country of origin than the MAM, and the more so the more strongly origin and sex are consolidated. Fig.~\ref{fig:data}A confirms this. There, consolidation is described by the absolute value of the normalized consolidation~$\kappa=(k - \fB)/\fB$, the distance of~$k$ from independence relative to~$\fB$, normalized so that it equals~$-1$ when no member of~$A_0$ belongs to~$B_0$, whatever the shares (Materials and Methods). In classes where the two attributes are close to independent, the two models estimate similar preferences over origin; in strongly consolidated classes, the SAM estimates a stronger one. Unable to control for preference over sex, the SAM attributes to origin the segregation that the preference over sex induces on it.

The size of the inflation has a direct reading. In a model with a single attribute the index equals the preference (the first case of Eq.~\ref{eq:chiA_minus_alpha}), and the SAM is fitted to the ties within and between origin groups, so its preference approximates the index of origin measured on the network, and equals it when the two origin groups have the same size. The MAM splits the same index into the direct preference over origin and the part induced by the preference over sex. The difference between the two estimated preferences, plotted in Fig.~\ref{fig:data}A, is therefore an estimate of the segregation on origin induced by sex in each class. Fig.~\ref{fig:data}B compares it with the induced segregation predicted by Eq.~\ref{eq:chiA_minus_alpha} from the composition of each class and the preferences estimated with the MAM. In most classes the difference lies close to this prediction; three classes, with predicted induced segregation between about 0.08 and 0.13, lie well below it.

Networks simulated from the two fitted models give the same picture: the SAM reproduces the segregation by the attribute it is fitted to, in this case country of origin, but produces almost no segregation by sex, whereas the MAM reproduces both (SI Appendix, Fig.~\ref{S-fig:replicas_origin}). The SAM fitted on sex is discussed in SI Appendix, section~\ref{S-si:replicas}.

\subsection{Splitting a school into two classes}
Following Moody \citep{moody2001race}, who argues that segregation is more realistically addressed by shaping opportunities for contact than by directly changing preferences, we consider a school administrator, whom we call the planner, who chooses the composition of the classes, and we ask what reduction in segregation they can achieve by splitting a school into two classes. 
We take segregation itself as the planner's objective, as in desegregation policy; we do not model the educational outcomes through which segregation matters. This subsection defines the splits and the two objectives with which the planner judges them.

The two classes together contain all the students of the school, so the composition of the school is conserved (Materials and Methods), and the ties within each class follow the MAM with the same preferences $\alpha$ and~$\beta$ as the whole school. The planner judges a split by the averages of the class indices weighted by class size, $\langle\chiA\rangle$ and $\langle\chiB\rangle$; the weighted average is the expected segregation of the class of a student drawn at random from the whole school. We consider two objectives,
\begin{equation}\label{eq:J}
J = \bigl|\langle\chiA\rangle\bigr| + \bigl|\langle\chiB\rangle\bigr|, \qquad
J_{\rm strict} = \bigl\langle|\chiA|\bigr\rangle + \bigl\langle|\chiB|\bigr\rangle \;\geq\; J ,
\end{equation}
where $\langle x\rangle=\sum_c (N^c/N)\,x^c$ is the average over the classes~$c$, weighted by their numbers of students~$N^c$. Both take absolute values, so that an excess and a deficit of within-group ties count alike as departures from the baseline. The aggregate objective~$J$ takes the absolute value of each average, so classes whose indices have opposite signs offset each other in it. The strict objective~$J_{\rm strict}$ takes absolute values class by class: it judges each class by its own segregation, and it is never smaller than~$J$. 

Eq.~\ref{eq:chiA_minus_alpha} shows which part of the segregation on~$A$ a planner can act on. Its first factor, the preference over~$B$, is out of the planner's reach; the other two, the consolidation $k-\fB$ and the difference~$\Delta$ between the $B$-imbalances, are set by the class compositions.

\subsection{With one neutral preference, only induced segregation can be removed}
In this subsection we consider the case of no preference over~$A$ ($\alpha=0$), so that all the segregation on~$A$ is induced, and the closed form tells what the composition of the classes can and cannot change.

\begin{figure}[t]
\centering
\includegraphics[width=\textwidth]{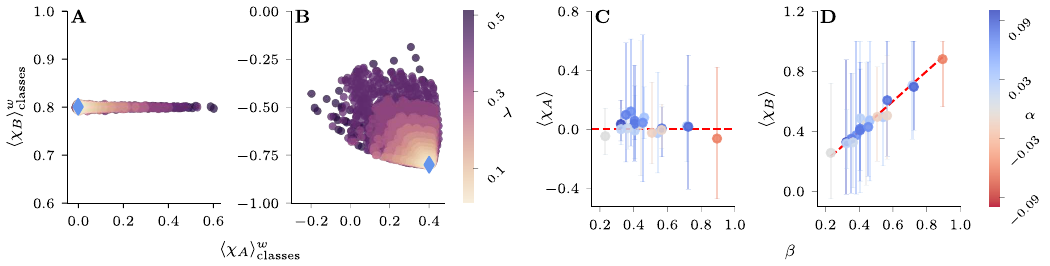}
\caption{\textbf{With no preference over $A$, class composition changes only the segregation on $A$.} (A and B) Size-weighted averages of the class indices, $\langle\chiA\rangle$ and $\langle\chiB\rangle$ (written $\langle\chi\rangle^{w}_{\rm classes}$ on the axes), measured on one simulated school ($\fA=0.2$, $\fB=0.5$, $k=0.5$, $N=800$) split into two classes in many ways; each point is a split, colored by the relative size~$\lambda$ of class~1, and the diamond is the undivided school. (A)~No preference over~$A$ ($\alpha=0$, $\beta=0.8$): $\langle\chiB\rangle$ stays close to~$\beta$, while $\langle\chiA\rangle$ ranges from about 0 to 0.6. (B)~Both preferences active ($\alpha=0.4$, $\beta=-0.8$): both averages vary. (C and D) ESDS splits of the 19 CILS4EU classes with $|\alpha|\leq0.1$ that lie in the region where the ESDS exists: (C)~$\langle\chiA\rangle$ and (D)~$\langle\chiB\rangle$ against the estimated~$\beta$, colored by the estimated~$\alpha$; dashed lines: the predictions $0$ and~$\beta$. The aggregate objective~$J$ of these splits is in SI Appendix, Fig.~\ref{S-fig:esds_J}.}
\label{fig:alpha0}
\end{figure}

If we consider Eq.~\ref{eq:chiA_minus_alpha} exchanging the role of the two attributes, the level of $B$-segregation is fixed by the preference over~$B$:
\begin{equation}\label{eq:chiB_half}
\chiB = \beta \qquad (\alpha = 0,\; \beta > 0)
\end{equation}
in every class, whatever its composition, provided that the preference over~$B$ is homophilic. %
With $\alpha=0$, attribute~$A$ contributes the same factor to every tie, so from the standpoint of~$B$ the model is a single-attribute model, in which the index equals the preference. With heterophily over~$B$ ($\beta<0$), the index of~$B$ equals~$\beta$ only when the two $B$-groups have the same size (SI Appendix, section~\ref{S-si:chiB_alpha0}). Under homophily, therefore, no assignment of students to classes can change segregation on~$B$: this is a policy impossibility (SI Appendix, section~\ref{S-si:splits}). Fig.~\ref{fig:alpha0}A shows it on a simulated school split in many different ways: with $\alpha=0$, the average index of~$B$ stays close to $\beta=0.8$, while the average index of~$A$ ranges from about~0 to~0.6. When both preferences are active, both averages vary (Fig.~\ref{fig:alpha0}B).

Segregation on~$A$, instead, can be removed through class composition. At $\alpha=0$, Eq.~\ref{eq:chiA_minus_alpha} reduces to $\chiA=\beta\,\fA\,(k-\fB)\,\Delta/[\fA(1-\fA)+\beta\widetilde{Q}_A]$ in the homophilic regime, and the heterophilic regime has the same numerator (SI Appendix, section~\ref{S-si:numA_alpha0}). The index of~$A$ therefore vanishes in a class in which the two attributes are independent ($k=\fB$) or in which $\Delta=0$. To remove segregation on~$A$ from every class, the planner can impose only the first condition on all of them. The second cannot hold in every class: $\Delta$ adds and subtracts fractions of the population (Materials and Methods), and each fraction of the school is the size-weighted average of the same fraction in its classes (Eq.~\ref{eq:conservation}), so the $\Delta$ of the school is the size-weighted average of the $\Delta$ of its classes, and when it is not zero, at least one class has $\Delta\neq0$. Independence, instead, can hold in every class: the classes can differ in composition so that the correlation between the two attributes in the school comes entirely from the differences between the classes, while within each class the two attributes are independent. We do not consider the $B$-stratified split, which puts all the students of~$B_0$ in one class and all those of~$B_1$ in the other: it makes the index of~$A$ zero only by separating the two $B$-groups completely, which leaves the index of~$B$ undefined.

\begin{figure}[t]
\centering
\includegraphics[width=0.6\textwidth]{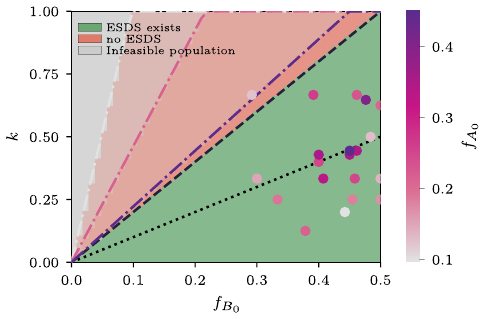}
\caption{\textbf{The decorrelating split exists in 19 of the 20 classes with $\alpha\approx0$.} Plane of the share of~$B_0$ in the school, $\fB$, and the consolidation~$k$. An equal-size decorrelating split (ESDS) exists below the line $k=2\fB$ (dashed; green region) and does not exist above it, where the population is feasible (red region); gray marks infeasible populations. Dash-dotted lines: the population feasibility ceiling $k=\min(1,\fB/\fA)$ for $\fA=0.1$, $0.21$ and~$0.45$, colored as in the color bar. Dotted line: independence, $k=\fB$. At each grid point, an ESDS is searched for numerically within the one-parameter family of SI Appendix, section~\ref{S-si:esds}. Points: the 20 CILS4EU classes with $|\alpha|\leq0.1$, colored by~$\fA$; all but one lie in the region where the ESDS exists.}
\label{fig:esdsmap}
\end{figure}

A split that makes the two attributes independent within each class, while keeping both groups of each attribute in every class, is what we call a decorrelating split. Independence within each class, together with the conservation of the composition of the school (Eq.~\ref{eq:conservation}), requires, for two classes of equal size,
\begin{equation}\label{eq:esds}
\bigl(\fA^{c_1}-\fA\bigr)\bigl(\fB^{c_1}-\fB\bigr)=\fA\,(k-\fB),
\end{equation}
where $\fA^{c_1}$ and $\fB^{c_1}$ are the shares of~$A_0$ and~$B_0$ in one of the two classes (SI Appendix, section~\ref{S-si:esds}). The right-hand side is the covariance of the two attributes in the school: the fraction of students who belong to both $A_0$ and~$B_0$, minus its value $\fA\fB$ under independence. The left-hand side is the covariance created by the difference between the two classes. Since in a decorrelating split the attributes are independent within each class, the correlation of the school must be carried entirely by the difference between the classes. A class with half of the students contains at most twice the school's share of each group, so the left-hand side cannot exceed $\fA\fB$ in absolute value, and Eq.~\ref{eq:esds} has solutions if and only if $k\leq2\fB$, that is, unless the fraction of~$A_0$ students who belong to~$B_0$ exceeds twice the fraction of~$B_0$ students in the school. 
The solutions to the problem of how to split a school in two equal-sized classes in which the two attributes within each class are decorrelated form a one-parameter family (Materials and Methods); we call them equal-size decorrelating splits (ESDS). Fig.~\ref{fig:esdsmap} shows where they exist, in the plane of the share~$\fB$ and the consolidation~$k$. When $k>2\fB$, no split into two equal classes makes the attributes independent in both; this is the case for one of the 20 classes of our data with $\alpha\approx0$ (Fig.~\ref{fig:esdsmap}).

We thus established that segregation on~$B$ equals~$\beta$ in every class, and segregation on~$A$ can be brought to zero, so the minimum of~$J$ is~$\beta$: an irreducible floor (SI Appendix, section~\ref{S-si:optimum_alpha0}). When $0<k<2\fB$, an ESDS attains it while keeping both groups of both attributes in each class. Since every class then has $\chiA=0$ and $\chiB=\beta$, the ESDS also gives $J_{\rm strict}=J=\beta$, and it is optimal for both objectives. The best a school can do is to remove all the induced segregation on~$A$; segregation on~$B$ stays at the level set by the preference over~$B$.

A careless split can do the opposite. In the school of Fig.~\ref{fig:alpha0}A, the two attributes are independent, so the undivided school has no segregation on~$A$. Yet many splits give classes with a positive average index of~$A$, up to about~0.6: a split can generate racial segregation where there was none.

We test these predictions on the 20 classes, among the 69 in which the MAM fits as well as the uMAM, whose estimated preference over~$A$ is close to zero ($|\alpha|\leq0.1$). As shown in Fig.~\ref{fig:esdsmap}, all but one of them lie in the region where the ESDS exists. Since each network is a single class, we split each of these 19 classes into two subclasses with the compositions prescribed by the ESDS, keeping the ties observed within each subclass (Materials and Methods). Across the classes, the average index of~$A$ is centered on zero and the average index of~$B$ follows~$\beta$ (Fig.~\ref{fig:alpha0}C--D), so that $J$ lies close to the floor~$\beta$, within the error bars (SI Appendix, Fig.~\ref{S-fig:esds_J}). These splits keep the friendships observed in the undivided class, so they test the model on subsets of students; they do not measure the effect of an actual reassignment, after which students may form friendships differently \citep{carrell2013natural}. We compare the ESDS with two other splitting recipes. A split that correlates the two attributes performs worse than the ESDS (SI Appendix, section~\ref{S-si:splittests}), but a random split performs almost as well in these data. This result can be explained by the level of attribute consolidation found in the CILS4EU classes. On simulated schools, we find that the advantage of the ESDS over a random split grows with consolidation (SI Appendix, section~\ref{S-si:esds_numerics}), and the attributes in our data happen to be only moderately consolidated, making the performance of the ESDS similar to that of a random split.

\subsection{With two active preferences, splitting helps only when they conflict}
When both preferences are active, the planner faces a trade-off: a split that lowers segregation on one attribute may raise it on the other (Fig.~\ref{fig:alpha0}B). No closed-form solution is known for a general school. We therefore study a school balanced on both attributes, $\fA=\fB=1/2$. It is tractable, and it is the most diverse on both attributes: its consolidation can range from perfect anti-correlation to perfect correlation, symmetrically around independence, so that the interplay between the two preferences is most visible. This subsection gives the indices of a balanced school and solves the planner's problem when the two attributes are independent.

In a balanced school, the two $B$-imbalances are $\delta_0=-\delta_1=(2k-1)/2$, so the closed form, Eqs.~\ref{eq:chiA_closedform} and~\ref{eq:QA}, depends on the composition only through the squared consolidation $s=(2k-1)^2$, which is 0 at independence and 1 when the two attributes are perfectly correlated or anti-correlated. The indices of the two attributes become (SI Appendix, section~\ref{S-si:C12_index})
\begin{equation}\label{eq:chiAB_C3_s}
\chiA(s) = \frac{\alpha + \beta s}{1 + \alpha\beta s}, \qquad
\chiB(s) = \frac{\beta + \alpha s}{1 + \alpha\beta s}.
\end{equation}
The same formula holds in both the homophilic and the heterophilic regimes, because in a balanced school the two regimes have the same denominator (Materials and Methods). In Eq.~\ref{eq:chiAB_C3_s} consolidation controls how strongly the indices of the two attributes, and thus their levels of segregation, are coupled. At independence ($s=0$), each index equals its own preference. As consolidation grows, each index takes up more of the other preference, and under perfect consolidation ($s=1$) both take the same value $(\alpha+\beta)/(1+\alpha\beta)$. The induced segregation of Eq.~\ref{eq:chiA_minus_alpha} becomes $\chiA-\alpha=\beta(1-\alpha^2)\,s/(1+\alpha\beta s)$: it is zero at independence, has the sign of~$\beta$, and grows in absolute value with consolidation. Without preference over~$A$, $\chiA=\beta s$ is all induced, negligible near independence and comparable to~$\beta$ under strong consolidation.

The direction in which each index moves as consolidation grows is set by the other preference: by Eq.~\ref{eq:chiAB_C3_s}, the slope of~$\chiA$ with respect to~$s$ has the sign of~$\beta$, and that of~$\chiB$ the sign of~$\alpha$ (SI Appendix, section~\ref{S-si:C12_index}). We call agreement phase the case in which the two preferences have the same sign ($\alpha\beta>0$), and conflict phase the case in which they have opposite signs ($\alpha\beta<0$). In the conflict phase, labeling the attributes so that $|\alpha|\leq|\beta|$, the index of~$A$ vanishes at the squared consolidation $s^*=|\alpha|/|\beta|$, which we call the cancellation point.

\begin{figure}[t]
\centering
\includegraphics[width=\textwidth]{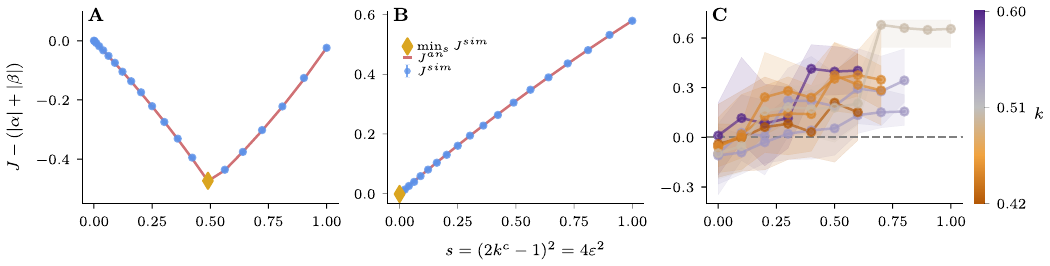}
\caption{\textbf{Splitting a balanced school at independence lowers segregation only when the two preferences conflict.} Aggregate objective~$J$ minus its value without a split, $|\alpha|+|\beta|$, against the squared consolidation $s=4\varepsilon^2$ of the two classes, for equal-size splits of a balanced school at independence ($\fA=\fB=1/2$, $k=1/2$) that keep its shares. (A and B) Calculated from the closed form (line, $J^{an}$ in the legend) and measured on simulated networks (symbols, $J^{sim}$; each class a network of 500 individuals, 30 networks per point). (A)~Conflict phase ($\alpha=0.4$, $\beta=-0.8$): without a split $J=|\alpha|+|\beta|=1.2$ and by splitting it can be lowered to its minimum $5/7\approx0.71$ at $s^*=1/2$ (diamond, $\min_s J^{sim}$ in the legend), where segregation on~$A$ vanishes in both classes. Changing the signs of both preferences leaves~$J$ unchanged, so the same curve holds for $\alpha=-0.4$ and $\beta=0.8$, heterophily on~$A$ and homophily on~$B$. (B)~Agreement phase ($\alpha=\beta=0.4$): $J$ grows with~$s$ and the best option is not to split. (C) Validation on data: $J-(|\alpha|+|\beta|)$ against the squared offset $4\varepsilon^2$ for the 7 CILS4EU classes close to the balanced school and to independence, all in the agreement phase; color:~$k$. As predicted, $J$ grows with the offset.}
\label{fig:C12}
\end{figure}

Consider a school at independence ($k=1/2$), split into two classes of equal size that both keep the population shares of the school. Each class is then balanced as well, and the only freedom left is its consolidation: by the conservation of the school's composition, the two classes have consolidations $1/2\pm\varepsilon$, for an offset~$\varepsilon$ between 0 and~1/2 (Materials and Methods). This makes their correlations between the attributes opposite, but their squared consolidations  the same, $s=4\varepsilon^2$. Since $s$ is the same between the two classes and the two classes have a balanced population, the two classes have the same index values given by Eq.~\ref{eq:chiAB_C3_s}. Averaging and taking absolute values then commute, so that $J_{\rm strict}=J=|\chiA(s)|+|\chiB(s)|$, and the planner chooses a single number,~$s$. At $s=0$ both classes inherit the composition of the school; since the index depends on the composition of a class and not on its size, having squared consolidation $s=0$ is the same as not splitting. In the agreement phase, $|\chiA|$ and $|\chiB|$ both grow with~$s$ (Eq.~\ref{eq:chiAB_C3_s}), so $J$ grows too, and the best option is not to split. In the conflict phase, $J$ falls as $s$ grows up to~$s^*$, where segregation on~$A$ vanishes in both classes, and rises beyond it (SI Appendix, section~\ref{S-si:C12_transition}). Its minimum,
\begin{equation}\label{eq:Jmin}
J_{\min}=\frac{\beta^2-\alpha^2}{|\beta|\,(1-\alpha^2)},
\end{equation}
lies below the floor $|\beta|$ found without preference over~$A$. Fig.~\ref{fig:C12}A--B shows the two phases on simulated networks.

The sign of $\alpha\beta$ matters because a split changes segregation only through the consolidation of the classes. Consider students who mildly avoid friends of the same race ($\alpha<0$) but prefer friends of their own gender ($\beta>0$); for country of origin and sex, this sign pattern occurs in some classes of our data. %
In a class where race and gender are correlated, gender homophily induces segregation by race, which offsets the direct heterophily by race; at the consolidation~$s^*$ the two cancel, and segregation by race vanishes. For $\alpha=-0.2$ and $\beta=0.6$, $J$ drops from $|\alpha|+|\beta|=0.80$ in the undivided school to~$0.56$. When the two preferences have the same sign, the induced and the direct effects add up instead, and any split increases segregation. Geometrically, a split at independence moves both classes to the same point of the segment traced by the two indices, and it can lower~$J$ only if the segment crosses an axis, which happens only in the conflict phase.

The recommendation not to split in the agreement phase, or equivalently to give both classes the composition of the school, holds away from independence as well, up to gains of at most $0.12\%$ of~$J$ in strongly consolidated schools (SI Appendix, section~\ref{S-si:C12_transition}).
 The estimated signs of the two preferences therefore tell the planner, before any split is designed, whether splitting can help.

We test the predictions at independence on the 7 classes, among the 69, that are close to the balanced school and to independence ($0.4\leq\fA,\fB\leq0.5$ and $|k-1/2|\leq0.1$; Materials and Methods). All of them are in the agreement phase, with $\alpha>0$ and $\beta>0$, and the result we just found predicts that the best strategy is to not split as any offset should increase~$J$. We split each class into two subclasses with consolidations $k\pm\varepsilon$, keeping the ties observed within each subclass, as for the ESDS. Within the error bands, $J$ increases with the squared offset~$4\varepsilon^2$, as predicted (Fig.~\ref{fig:C12}C). Deviations from this prediction grow as $k$ departs from~$1/2$, where the two subclasses are no longer mirror images. No class of this sample is in the conflict phase, so the prediction that splitting then helps is not tested on the data.

\subsection{A school average can hide classes that are more segregated}
At independence, the two classes of an equal-size split are mirror images, and the school average describes both. Away from independence ($k\neq1/2$) the two classes differ, and the school average can mislead the planner. This subsection shows how this can happen in a school with a balanced population whose preference parameters are in the conflict phase, and how judging a split class by class avoids it.

With an offset~$\varepsilon$, the two classes of an equal-size split that keeps the population shares of the school have consolidations $k\pm\varepsilon$ and squared consolidations
\begin{equation}\label{eq:spm}
s_\pm=(2k-1\pm2\varepsilon)^2,
\end{equation}
which differ unless $k=1/2$. The school averages of their indices are $\langle\chiA\rangle=[\chiA(s_+)+\chiA(s_-)]/2$ and $\langle\chiB\rangle=[\chiB(s_+)+\chiB(s_-)]/2$.

\begin{figure}[t]
\centering
\includegraphics[]{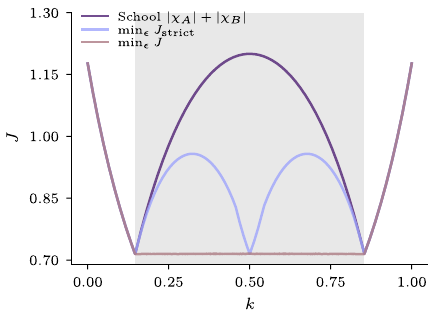}
\caption{\textbf{The school average can hide classes that are more segregated.} Balanced school ($\fA=\fB=1/2$) in the conflict phase ($\alpha=0.4$, $\beta=-0.8$, $s^*=1/2$), with equal-size splits that keep its shares; all curves are derived from the closed analytical forms of $\chiA$ and $\chiB$. Changing the signs of both preferences leaves $J$ and~$J_{\rm strict}$ unchanged, so the same curves hold for $\alpha=-0.4$ and $\beta=0.8$. Minimum over the offset of the aggregate objective~$J$ (rose) and of the strict objective~$J_{\rm strict}$ (lilac), and the undivided school (purple), against the school's consolidation~$k$. Inside the straddling window (gray band), $J$ reaches a constant floor by cancellation between classes, whereas $J_{\rm strict}$ reaches it only at $k=1/2$; outside the window the three curves coincide. For these parameters the canceling split exists at every~$k$ in the window; for given preferences this holds if and only if $s^*\leq(1+\alpha^2)/2$ (SI Appendix, section~\ref{S-si:C12_trap}).}
\label{fig:trap}
\end{figure}

In the conflict phase, a split can place the two classes on opposite sides of~$s^*$, so that segregation on~$A$ is positive in one class and negative in the other. Their average can then vanish, and $J$ reaches its floor, $|\chiB(s^*)|$, although no class has zero segregation on~$A$ and the zero is merely a cancellation between classes. Such a split exists only if the squared consolidation of the school is below the cancellation point, $(2k-1)^2<s^*$; for $\alpha=0.4$ and $\beta=-0.8$ this type of split exists for every such school%
, but this is not the case for a general pair of preferences (SI Appendix, section~\ref{S-si:C12_trap}). For schools far enough from independence (for $\alpha=0.4$ and $\beta=-0.8$, when $|2k-1|>0.30$), segregation on~$A$ is then larger, in absolute value, in every class than in the undivided school (SI Appendix, section~\ref{S-si:C12_trap}). Students experience the segregation of their own class, not the school average, so the reduction shown by~$J$ is only apparent. We call this the cancellation trap. Fig.~\ref{fig:trap} shows it for $\alpha=0.4$ and $\beta=-0.8$: for a range of consolidations, the minimum of~$J$ lies below what can be achieved class by class.

The remedy is to judge a split class by class, with $J_{\rm strict}$, which cannot be lowered by cancellations. For the splits of Eq.~\ref{eq:spm}, $J_{\rm strict}=[T(s_+)+T(s_-)]/2$, where $T(s)=|\chiA(s)|+|\chiB(s)|$ is the total segregation of a class. In the conflict phase, $T$ falls as $s$ grows up to~$s^*$ and rises beyond it (Materials and Methods). Judged in this way, a split helps only if the squared consolidation of the school is below the cancellation point,
\begin{equation}\label{eq:window}
(2k-1)^2<s^*,
\end{equation}
that is, if in the undivided school the direct preference over~$A$ still dominates the segregation induced by the preference over~$B$ (SI Appendix, section~\ref{S-si:C12_strict}). We call this range of consolidations, centered on independence, the straddling window, and it widens as the two preferences become comparable in strength. Outside the window, the school already lies on the rising side of~$T$, moving the two classes apart cannot lower their average, and the best option is not to split. Inside the window, the best split usually places one class at the cancellation point, where segregation on~$A$ vanishes, and the other class at the consolidation that the conservation of the school's composition then imposes (Materials and Methods); exceptions   are derived in SI Appendix, section~\ref{S-si:C12_strict}.%

Judging class by class has a cost: away from independence, the minimum of~$J_{\rm strict}$ lies above the floor of~$J$ (Fig.~\ref{fig:trap}), but the reduction it delivers is real, not an effect of cancellations between classes. %
Even so, in schools far enough from independence (for $\alpha=0.4$ and $\beta=-0.8$, when $|2k-1|>0.24$), the optimal split leaves the second class more segregated than the undivided school on both attributes (SI Appendix, section~\ref{S-si:C12_both}).

The trap is therefore confined to schools with opposed preferences whose consolidation lies inside the straddling window. In the conflict phase and outside the window, $J$ and $J_{\rm strict}$ have the same minimum, obtained by not splitting (SI Appendix, section~\ref{S-si:C12_strict}). In the agreement phase, neither index changes sign (Eq.~\ref{eq:chiAB_C3_s}), so each index has the same sign in every class, averaging and taking absolute values commute, $J=J_{\rm strict}$ for every split, and the results previously found for $J$ apply also to $J_{\rm strict}$.

\section{Discussion}
Segregation measured on one attribute is usually read as the result of the preference over that attribute and of the opportunities for contact.
 Our results show when this reading fails and what follows for the composition of classes. We summarize them, relate them to previous work, and discuss their limitations and possible extensions.

We have calculated in closed form the Coleman index of each attribute in a network model in which ties depend on preferences over two correlated attributes. Segregation measured on one attribute combines the preference over that attribute with segregation induced by the preference over the other, so a single-attribute reading can attribute it to the wrong cause. In school friendship networks, a model with preferences over both country of origin and sex describes the ties better than a model with origin alone in every class, and the latter inflates the preference over origin where origin and sex are consolidated. When the preference over one attribute is neutral, a planner who assigns students to classes can remove the induced segregation entirely, but cannot change segregation on the other attribute, which, under homophily, equals its preference in every class. In a school balanced on both attributes, with the two attributes independent, splitting can reduce segregation only when the two preferences have opposite signs. Away from independence, judging a split by the school average can hide classes that are each more segregated on one attribute than the undivided school. The splitting predictions hold on the classes of our data to which they apply.

For a school, the results translate into three points. First, the signs of the two preferences, estimated from the friendship networks of existing classes, tell in advance whether splitting can help: it cannot when the preferences agree, and it can when they conflict and the consolidation of the school lies inside the straddling window. Second, a split should be judged by the segregation inside each class, not by the school average; the two criteria agree when the preferences agree and outside the straddling window, so the cancellation trap is confined to schools with opposed preferences whose consolidation lies inside the window. Third, our results on the optimal split assume a school balanced on both attributes and two classes of equal size that keep its shares, but the inequality $J\leq J_{\rm strict}$, and with it the reason to judge a split class by class, holds for any school and any split.

Currarini, Jackson and Pin \citep{currarini2009economic, currarini2010identifying} provided an economic model of homophily in friendship networks, estimating preferences on the Add Health data \citep{harris2013add} within an equilibrium framework in which meeting rates and the degree distribution are determined jointly. The reduced-form model we adopt instead takes meeting rates as given and extends type-dependent acceptance probabilities from one attribute to two correlated ones. With two attributes, induced segregation adds a third source to the two forces through which segregation measured on a single attribute is usually read: preferences for similar others, and the opportunities for contact set by group sizes \citep{blau1977inequality, currarini2009economic, currarini2010identifying}. This gives an analytical explanation of the finding that friendship segregation is stronger in classrooms where ethnicity and gender are aligned \citep{kroneberg2021ethnicity, boller2025friendship}. In the MAM, induced segregation requires both a preference over the second attribute and a difference between the two groups in their composition on it, which matches the prediction of Kroneberg and Lorenz \citep{kroneberg2024structural} that consolidation matters only when homophily on the second attribute is strong. In their formalization of Blau's theory, however, ties arise from homophily on either attribute or from chance, rather than from a joint evaluation of both attributes as in the MAM. Unlike them, we calculate a normalized segregation index and derive class-placement strategies that reduce segregation.

Our work also sits between two recent papers by Martin-Gutierrez, Cartier van Dissel and Karimi. They compare several models of intersectionality and identify the multiplicative model as the one that best reproduces empirical networks \citep{martin2026simple}, and we adopt this model. They also study the effects of intersectionality on social capital \citep{martin2025intersectional}, whereas we focus on segregation as an inequality-driving phenomenon.

On the policy side, our work builds on Moody's intuition that school administrators have more leverage over the organization of the school than over the behavior of students \citep{moody2001race}, and it makes the limits of this approach to desegregation explicit. Assuming that preferences do not depend on the opportunities for contact, class composition can remove the segregation that one preference induces on the other attribute, but it cannot change segregation set by a homophilic preference over the attribute itself. When both preferences are active, in a balanced school with independent attributes, class composition can reduce segregation only if the two preferences have opposite signs. B\"oller and Kruse \citep{boller2025friendship} evaluate class-assignment heuristics by simulation; our closed forms show analytically when splitting a school with two attributes reduces segregation, when it increases it, and when it only appears to reduce it.

We identify five main limitations.
First, the planner's objective is segregation itself: we do not model how segregation affects educational or other outcomes, a question studied by a large empirical literature \citep{armor1995forced, braddock1988social, wells1994perpetuation, mickelson2001subverting}. Second, the model is static, and its ties are independent of each other. It predicts friendships for a given class composition, but after an actual reassignment students may form ties differently than predicted \citep{carrell2013natural}, and it rules out triangle closure and other effects of existing friendships on new ones; our tests split existing classes and keep their friendships, so they check the model, not the effect of a reassignment. Third, the attributes are binary, each preference is the same for both groups of an attribute, and the two preferences combine multiplicatively. The network density and the Coleman index keep closed forms with more groups per attribute or with group-specific preferences, at the cost of less readable expressions, but the simplifications of the two cases we solve do not automatically carry over. Fourth, the estimated preferences are only as credible as the model that defines them. Identification is not the issue: since ties are independent given the types of the two individuals, the preferences follow from the counts of ties between types. The challenge is model selection, and our comparisons with the single-attribute and unconstrained models, together with the evidence that the multiplicative model reproduces school friendship networks better than a non-factorized one \citep{martin2026simple}, support the multiplicative form. Fifth, the data test only part of the predictions: the balanced classes close to independence are all in the agreement phase, so the conflict phase, in which splitting reduces segregation, and the cancellation trap remain untested on data.

Two extensions follow naturally. The multiplicative model could be embedded in a search-and-matching equilibrium with two attributes, in the style of Currarini, Jackson and Pin \citep{currarini2009economic}, in which the preferences would depend on the composition of the population. The planner's problem could also include incentive constraints: if students can choose their class, the optimal partition must be incentive-compatible, and the problem becomes one of mechanism design with segregation externalities.

\section{Materials and Methods}
\subsection{Model}
This subsection specifies the population composition and the MAM preference matrices described in the Results; their properties are given in SI Appendix, section~\ref{S-si:model}.
The fraction $F_{A_a,B_b}$ of individuals of each type follows from the shares $\fA$, $\fB$ and the consolidation~$k$:
\begin{equation}\label{eq:Fparam}
\begin{aligned}
F_{A_0,B_0} &= k\,\fA, & F_{A_0,B_1} &= (1-k)\,\fA,\\
F_{A_1,B_0} &= \fB - k\,\fA, & F_{A_1,B_1} &= 1 - \fB - (1-k)\fA .
\end{aligned}
\end{equation}
All four fractions must be non-negative. Under the minority convention of Martin-Gutierrez et al.\ \citep{martin2025intersectional}, which we adopt, $\fA,\fB\leq 1/2$, which restricts $k$ to the interval from $0$ to $\min(1,\fB/\fA)$; at the two ends of this interval the two attributes are maximally anti-correlated and maximally correlated. The $B$-imbalance of group~$A_a$ is $\delta_a=F_{A_a,B_0}-F_{A_a,B_1}$, so that
\begin{equation}\label{eq:delta}
\delta_0=(2k-1)\,\fA,\qquad \delta_1=(2\fB-1)-(2k-1)\,\fA .
\end{equation}
The two imbalances add up to the $B$-imbalance of the population, $\delta_0+\delta_1=2\fB-1$, and their difference is $\Delta=\delta_0-\delta_1=2\fA(2k-1)-(2\fB-1)$. The normalized consolidation $\kappa=(k-\fB)/\fB$ is zero at independence and equals $-1$ at $k=0$, whatever the shares; an equal-size decorrelating split exists if and only if $\kappa\leq1$ (see the Results Section).

In the original model \citep{martin2026simple}, called MAPS-AND by its authors, the two $2\times2$ preference matrices of Eq.~\ref{eq:link_prob} have free entries, eight in total if we consider two attributes, of which seven are free, since multiplying one matrix by a constant and dividing the other by it leaves every tie probability unchanged; we call this unconstrained version the uMAM. The MAM makes each matrix symmetric and normalized, with $h$ on the diagonal and $1-h$ off it, and writes it in terms of the centered preferences $\alpha=2h_A-1$ and $\beta=2h_B-1$, so that each attribute has a single preference parameter. The single-attribute model (SAM) keeps only one of the two terms of Eq.~\ref{eq:link_prob}. The tie probability is defined up to a constant, which sets the overall density of ties; it does not affect the Coleman index, which compares fractions of ties. In the data, density is estimated along with the preference parameters.%

\subsection{Coleman index in the model}
This subsection defines the quantities used to compute the closed form in the Results (Eqs.~\ref{eq:chiA_closedform}--\ref{eq:chiA_minus_alpha}); the derivation is given in SI Appendix, section~\ref{S-si:maci}.
The Coleman baseline of attribute~$A$ is
\begin{equation}\label{eq:omega_star}
w_A^*= \frac{1}{\sum_{i}\eta_i}\,\sum_{g \in \{A_0, A_1\}} \;\sum_{i \in G_g} \eta_i \,\frac{n_g - 1}{N - 1},
\end{equation}
where $G_g$ is the set of members of group~$g$, $n_g$ its size, and $\eta_i$ the out-degree of individual~$i$, the number of ties it sends: when partners are chosen at random, each tie of a member of~$G_g$ reaches one of the other $n_g-1$ members of the group with probability $(n_g-1)/(N-1)$. Because the tie probabilities of the MAM are symmetric, the index is the same whether the network is treated as directed or undirected.

In the model, the index is calculated from expected densities of ties. Up to the constant of the tie probability, the density of ties within group~$A_a$ and the density of all the ties sent by its members are
\begin{equation}\label{eq:densities}
\begin{aligned}
W_a&=\tfrac{1+\alpha}{4}\bigl(f_{A_a}^2+\beta\,\delta_a^2\bigr),\\
\Sigma_a&=W_a+\tfrac{1-\alpha}{4}\bigl(f_{A_a}f_{A_{\bar a}}+\beta\,\delta_a\delta_{\bar a}\bigr),
\end{aligned}
\end{equation}
where $\bar a$ denotes the other $A$-group and $f_{A_1}=1-\fA$ (SI Appendix, section~\ref{S-si:densities}). The density of ties within $A$-groups is $W=W_0+W_1$, and the network density is $Z=\Sigma_0+\Sigma_1$. For large~$N$, the Coleman baseline equals $w_A^*=\fA\,\phi_0+(1-\fA)\,\phi_1$, where the degree share $\phi_a=\Sigma_a/Z$ is the fraction of all ties sent by members of group~$A_a$: each tie leaves a member of~$A_a$ with probability~$\phi_a$ and, under the baseline, reaches a member of the same group with probability~$f_{A_a}$ (SI Appendix, section~\ref{S-si:baseline}). When all individuals have the same expected number of ties, $\phi_a=f_{A_a}$, and the baseline reduces to the random-mixing value $\fA^2+(1-\fA)^2$. On the scale of the densities, the baseline is $W^*=Z\,w_A^*=\fA\,\Sigma_0+(1-\fA)\,\Sigma_1$.

The index is $\chiA=(W-W^*)/(Z-W^*)$ in the homophilic regime and $\chiA=(W-W^*)/W^*$ in the heterophilic regime. Substituting the densities gives Eqs.~\ref{eq:chiA_closedform} and~\ref{eq:QA}, in which the numerator is $2(W-W^*)$ and the denominator $2(Z-W^*)$. In the heterophilic regime,
\begin{equation}\label{eq:hetero}
\chiA=\chi_A^{\rm hom}\,\frac{Z-W^*}{W^*},
\end{equation}
where $\chi_A^{\rm hom}$ is the right-hand side of Eq.~\ref{eq:chiA_closedform} (SI Appendix, section~\ref{S-si:closedform}). When the two $A$-groups have the same size ($\fA=1/2$), $w_A^*=(\phi_0+\phi_1)/2=1/2$ whatever the degree shares, since $\phi_0+\phi_1=1$; hence $W^*=Z/2=Z-W^*$: the two regimes have the same denominator, and Eq.~\ref{eq:chiAB_C3_s} holds in both. The index of~$B$ follows from Eqs.~\ref{eq:chiA_closedform} and~\ref{eq:QA} by exchanging the roles of the two attributes: $\alpha$ with~$\beta$, $\fA$ with~$\fB$, and the $B$-imbalances of the $A$-groups with the $A$-imbalances of the $B$-groups, $\delta'_0=2k\fA-\fB$ and $\delta'_1=2(1-k)\fA-(1-\fB)$ (SI Appendix, section~\ref{S-si:attrB}).

\subsection{Two-class splits}
In this subsection we define the splits used in the Results and give the formulas for the optimal splits (derived in SI Appendix, sections~\ref{S-si:alpha0} and~\ref{S-si:C12}).
A split divides the $N$ students of a school into two classes $c_1$ and~$c_2$, with $N^{c_1}$ and~$N^{c_2}$ students and population matrices $F^{c_1}$ and~$F^{c_2}$. Since every student goes to one class,
\begin{equation}\label{eq:conservation}
N\,F=N^{c_1}F^{c_1}+N^{c_2}F^{c_2},
\end{equation}
and the averages of the class indices are $\langle\chi_X\rangle=\sum_c(N^c/N)\,\chi_X^c$, for $X=A,B$.

\textbf{ESDS} For two classes of equal size, Eq.~\ref{eq:esds} has a one-parameter family of solutions, the decorrelating splits; both groups of both attributes are present in each class when $0<k<2\fB$ (SI Appendix, section~\ref{S-si:esds}).

\textbf{Splits in a balanced school} In a balanced school, the splits of the Results have two classes of equal size that both keep the shares $\fA=\fB=1/2$. Eq.~\ref{eq:conservation} then leaves one free parameter, the offset~$\varepsilon$: the classes have consolidations $k\pm\varepsilon$, with $0\leq\varepsilon\leq\min(k,1-k)$, and the squared consolidations of Eq.~\ref{eq:spm}. In the conflict phase, take $\alpha>0>\beta$ with $|\alpha|\leq|\beta|$; the case $\alpha<0<\beta$ follows by changing the signs of both preferences, which leaves the absolute values of the indices unchanged. Eq.~\ref{eq:chiAB_C3_s} gives the total segregation of a class,
\begin{equation}\label{eq:Tbranches}
T(s)=
\begin{cases}
\dfrac{(\alpha-\beta)(1-s)}{1+\alpha\beta s}, & s\leq s^*,\\[8pt]
\dfrac{-(\alpha+\beta)(1+s)}{1+\alpha\beta s}, & s\geq s^*,
\end{cases}
\end{equation}
which decreases up to~$s^*$ and increases beyond it (SI Appendix, section~\ref{S-si:C12_strict}). The offsets that minimize $J_{\rm strict}$ inside the straddling window are derived in SI Appendix, section~\ref{S-si:C12_strict}.

\subsection{Simulations}
This subsection describes how networks are simulated from the MAM, to check the closed forms. A network of $N$ individuals contains $\mathrm{round}(F_{A_a,B_b}N)$ individuals of each type $(A_a,B_b)$, and any rounding discrepancy is assigned to the largest type. Each ordered pair of distinct individuals is connected by a directed tie, independently of all other pairs, with probability $H^A_{A(i),A(j)}H^B_{B(i),B(j)}$, as in Eq.~\ref{eq:link_prob}. Only weakly connected networks are kept: any other network is drawn again. The index and the density are measured on each simulated network and averaged over 30 networks of $N=2{,}000$ individuals per parameter value (SI Appendix, Figs.~\ref{S-fig:Z} and~\ref{S-fig:validation}). For Fig.~\ref{fig:C12}A--B, the two classes of each split are simulated as separate networks of 500 individuals, 30 times for each offset, and $J$ is computed from the indices measured on them.

\subsection{Data}
This subsection describes how the networks of the Results are built from the survey. We use the second wave of CILS4EU \citep{kalter2016cils4eu}, collected in 2011 and 2012. Students named classmates in several items on friendship and shared activities; each nomination gives a directed tie from the respondent to the classmate named, and a pair named in several items gives a single tie. We build one network per class, or per school in the few schools where nominations cross class boundaries. A student is of foreign origin if the student or either parent was born outside the survey country, and of survey-country origin otherwise \citep{dollmann2016examining, boller2025friendship}. We remove students with a missing attribute and keep the networks with at least 20 students, with at least one student of each of the four types $(A_a,B_b)$, and with a single connected component. This results in a total of 139 classes. SI Appendix, section~\ref{S-si:cils4eu}, lists the survey items used to build the networks and their attributes and summarizes the networks' properties by country (Table~\ref{S-tab:allnets}).

\subsection{Model estimation}
This subsection describes how the preferences are estimated from data and how the models are compared. For each class, we estimate the preferences by maximizing the binomial log-likelihood of the numbers of directed ties between the four types $(A_a,B_b)$, given the number of students of each type, with a method close to that of Martin-Gutierrez et al.\ \citep{martin2026simple}; a density parameter, estimated together with the preferences, lets the MAM and the SAM match the overall density of the class, as the free entries of the uMAM do. We fit three nested models, the MAM, the SAM and the uMAM, and compare each nested pair with a likelihood-ratio test at the 5\% significance level. The analyses that rely on the estimated preferences use the 69 classes in which the uMAM does not fit significantly better than the MAM. The likelihood, the density parameter and the tests are given in SI Appendix, section~\ref{S-si:inference}.

\subsection{Class-splitting experiments}
This subsection describes how the splitting results are tested on the data. Each empirical network is a single class, so we split classes into two subclasses rather than schools into classes: the students of each type are assigned to the subclasses at random, according to the target composition, 20 times per split, and only the ties within each subclass are kept. The ESDS test uses the 19 classes with $|\alpha|\leq0.1$ in which the ESDS exists; the balanced-school test uses the 7 classes with $0.4\leq\fA,\fB\leq0.5$ and $|k-1/2|\leq0.1$, split into subclasses with consolidations $k\pm\varepsilon$ that keep the shares of the class. The procedure is given in SI Appendix, section~\ref{S-si:splittests}.

\section*{Declarations}
\paragraph{Funding.} M.S. acknowledges support from Grants No.~RYC2022-037932-I and CNS2023-144156 funded by MCIN/AEI/10.13039/501100011033 and the European Union NextGenerationEU/PRTR.
\paragraph{Author contributions.} E.O. and M.S. designed research; C.E. performed research; C.E. analyzed data; and C.E., E.O., and M.S. wrote the paper.
\paragraph{Acknowledgements.}The authors wish to thank Alberto Bisin and Fabrizio Germano for their thoughtful comments and feedbacks on earlier versions of the manuscript.
\paragraph{Competing interests.} The authors declare no competing interest.
\paragraph{Data, materials, and software availability.} The CILS4EU data are third-party data, available for scientific use from the GESIS Data Archive for the Social Sciences after registration and signing a data use agreement~\citep{kalter2016cils4eu}. The authors do not have permission to redistribute the data directly. The code that computes the closed forms, simulates the networks, estimates the models and runs the class-splitting experiments will be made publicly available in Zenodo upon acceptance.
\paragraph{Use of generative AI.} We used the large language model Claude (Opus 5.5) to edit parts of the text and to check derivations of results in the SI Appendix. We checked all its outputs and take full responsibility for the content of the paper.

\bibliographystyle{unsrtnat}
\bibliography{references}

@article{currarini2009economic,
  title={An economic model of friendship: Homophily, minorities, and segregation},
  author={Currarini, Sergio and Jackson, Matthew O and Pin, Paolo},
  journal={Econometrica},
  volume={77},
  number={4},
  pages={1003--1045},
  year={2009},
  publisher={Wiley Online Library}
}

@article{currarini2010identifying,
  title={Identifying the roles of race-based choice and chance in high school friendship network formation},
  author={Currarini, Sergio and Jackson, Matthew O and Pin, Paolo},
  journal={Proceedings of the National Academy of Sciences},
  volume={107},
  number={11},
  pages={4857--4861},
  year={2010},
  publisher={National Academy of Sciences}
}

@article{jackson2025inequality,
  title={Inequality's economic and social roots: the role of social networks and homophily},
  author={Jackson, Matthew O},
  journal={arXiv preprint arXiv:2506.13016},
  year={2025}
}

@article{mcpherson2001birds,
  title={Birds of a feather: Homophily in social networks},
  author={McPherson, Miller and Smith-Lovin, Lynn and Cook, James M},
  journal={Annual Review of Sociology},
  volume={27},
  number={1},
  pages={415--444},
  year={2001},
  publisher={Annual Reviews 4139 El Camino Way, PO Box 10139, Palo Alto, CA 94303-0139, USA}
}

@article{moody2001race,
  title={Race, school integration, and friendship segregation in {America}},
  author={Moody, James},
  journal={American Journal of Sociology},
  volume={107},
  number={3},
  pages={679--716},
  year={2001},
  publisher={The University of Chicago Press}
}

@article{block2014multidimensional,
  title={Multidimensional homophily in friendship networks},
  author={Block, Per and Grund, Thomas},
  journal={Network Science},
  volume={2},
  number={2},
  pages={189--212},
  year={2014},
  publisher={Cambridge University Press}
}

@article{martin2025intersectional,
  title={Intersectional inequalities in social ties},
  author={Martin-Gutierrez, Samuel and {Cartier van Dissel}, Mauritz N and Karimi, Fariba},
  journal={Science Advances},
  volume={11},
  number={45},
  pages={eadu9025},
  year={2025},
  publisher={American Association for the Advancement of Science}
}

@article{coleman1958relational,
  title={Relational analysis: The study of social organizations with survey methods},
  author={Coleman, James S},
  journal={Human Organization},
  volume={17},
  number={4},
  pages={28--36},
  year={1958},
  publisher={Taylor \& Francis}
}

@article{bojanowski2014measuring,
  title={Measuring segregation in social networks},
  author={Bojanowski, Micha{\l} and Corten, Rense},
  journal={Social Networks},
  volume={39},
  pages={14--32},
  year={2014},
  publisher={Elsevier}
}

@article{card2007racial,
  title={Racial segregation and the black--white test score gap},
  author={Card, David and Rothstein, Jesse},
  journal={Journal of Public Economics},
  volume={91},
  number={11-12},
  pages={2158--2184},
  year={2007},
  publisher={Elsevier}
}

@article{chetty2016effects,
  title={The effects of exposure to better neighborhoods on children: New evidence from the moving to opportunity experiment},
  author={Chetty, Raj and Hendren, Nathaniel and Katz, Lawrence F},
  journal={American Economic Review},
  volume={106},
  number={4},
  pages={855--902},
  year={2016},
  publisher={American Economic Association 2014 Broadway, Suite 305, Nashville, TN 37203}
}

@article{chetty2022social,
  title={Social capital {I}: measurement and associations with economic mobility},
  author={Chetty, Raj and Jackson, Matthew O and Kuchler, Theresa and Stroebel, Johannes and Hendren, Nathaniel and Fluegge, Robert B and Gong, Sara and Gonzalez, Federico and Grondin, Armelle and Jacob, Matthew and others},
  journal={Nature},
  volume={608},
  number={7921},
  pages={108--121},
  year={2022},
  publisher={Nature Publishing Group UK London}
}

@article{martin2026simple,
  title={A simple preference aggregation rule explains how multidimensional identities shape social networks},
  author={Martin-Gutierrez, Samuel and {Cartier van Dissel}, Mauritz N and Karimi, Fariba},
  journal={Communications Physics},
  year={2026},
  doi={10.1038/s42005-026-02537-3},
  publisher={Nature Publishing Group UK London}
}

@article{kroneberg2021ethnicity,
  title={When ethnicity and gender align: Classroom composition, friendship segregation, and collective identities in {European} schools},
  author={Kroneberg, Clemens and Kruse, Hanno and Wimmer, Andreas},
  journal={European Sociological Review},
  volume={37},
  number={6},
  pages={918--934},
  year={2021},
  publisher={Oxford University Press}
}

@article{boller2025friendship,
  title={Friendship segregation and class composition in schools: A systematic analysis of the role of attribute consolidation},
  author={B{\"o}ller, Susanne and Kruse, Hanno},
  journal={PLoS One},
  volume={20},
  number={12},
  pages={e0339581},
  year={2025},
  publisher={Public Library of Science San Francisco, CA USA}
}

@book{armor1995forced,
  title={Forced justice: School desegregation and the law},
  author={Armor, David J},
  year={1995},
  publisher={Oxford University Press}
}

@article{braddock1988social,
  title={The social and academic consequences of school desegregation.},
  author={Braddock, Jomills Henry II and McPartland, James M},
  journal={Equity and Choice},
  volume={4},
  year={1988},
  publisher={ERIC}
}

@article{wells1994perpetuation,
  title={Perpetuation theory and the long-term effects of school desegregation},
  author={Wells, Amy Stuart and Crain, Robert L},
  journal={Review of Educational Research},
  volume={64},
  number={4},
  pages={531--555},
  year={1994},
  publisher={Sage Publications Sage CA: Thousand Oaks, CA}
}

@article{mickelson2001subverting,
  title={Subverting {Swann}: First- and second-generation segregation in the {Charlotte-Mecklenburg} schools},
  author={Mickelson, Roslyn Arlin},
  journal={American Educational Research Journal},
  volume={38},
  number={2},
  pages={215--252},
  year={2001},
  publisher={Sage Publications}
}

@article{dimaggio2012network,
  title={Network effects and social inequality},
  author={DiMaggio, Paul and Garip, Filiz},
  journal={Annual Review of Sociology},
  volume={38},
  number={1},
  pages={93--118},
  year={2012},
  publisher={Annual Reviews}
}

@article{boda2023ethnic,
  title={Ethnic diversity fosters the social integration of refugee students},
  author={Boda, Zs{\'o}fia and Lorenz, Georg and Jansen, Malte and Stanat, Petra and Edele, Aileen},
  journal={Nature Human Behaviour},
  volume={7},
  number={6},
  pages={881--891},
  year={2023},
  publisher={Nature Publishing Group UK London}
}

@article{killen2022reducing,
  title={Reducing prejudice through promoting cross-group friendships},
  author={Killen, Melanie and Luken Raz, Katherine and Graham, Sandra},
  journal={Review of General Psychology},
  volume={26},
  number={3},
  pages={361--376},
  year={2022},
  publisher={Sage Publications Sage CA: Los Angeles, CA}
}

@book{blau1977inequality,
  title={Inequality and heterogeneity: A primitive theory of social structure},
  author={Blau, Peter M.},
  year={1977},
  publisher={Free Press},
  address={New York}
}

@book{blau2018crosscutting,
  title={Crosscutting social circles: Testing a macrostructural theory of intergroup relations},
  author={Blau, Peter and Schwartz, Joseph},
  year={1984},
  publisher={Routledge}
}

@article{kroneberg2024structural,
  title={Structural sources of segregation: How classroom composition shapes interethnic friendships},
  author={Kroneberg, Clemens and Lorenz, Georg},
  journal={SocArXiv preprint},
  note={\url{https://doi.org/10.31235/osf.io/v4ne9}},
  year={2024}
}

@article{zhao2023superdiversity,
  title={From superdiversity to consolidation: Implications of structural intersectionality for interethnic friendships},
  author={Zhao, Linda},
  journal={American Journal of Sociology},
  volume={128},
  number={4},
  pages={1114--1157},
  year={2023},
  publisher={The University of Chicago Press Chicago, IL}
}

@book{rosche2024micro,
  title={Micro-Level Determinants of Macro-Level Outcomes. The Micro-Macro Link With Empirical Methods},
  author={Rosche, F Benjamin},
  year={2024},
  publisher={Cornell University}
}

@misc{kalter2016cils4eu,
  author       = {Kalter, Frank and Heath, Anthony F. and Hewstone, Miles and
                  Jonsson, Jan O. and Kalmijn, Matthijs and Kogan, Irena and
                  van Tubergen, Frank},
  title        = {{Children of Immigrants Longitudinal Survey in Four European Countries (CILS4EU)} -- {Reduced} version. {Reduced} data file for download and off-site use},
  howpublished = {GESIS Data Archive, Cologne. ZA5656 Data file Version 2.3.0. Available at \url{https://gesis.org}},
  year         = {2016}
}

@techreport{dollmann2016examining,
  title={Examining the diversity of youth in {Europe}: A classification of generations and ethnic origins using {CILS4EU} data},
  author={Dollmann, J{\"o}rg and Jacob, Konstanze and Kalter, Frank},
  institution={Mannheimer Zentrum f{\"u}r Europ{\"a}ische Sozialforschung (MZES)},
  type={Working paper},
  number={156},
  year={2014}
}

@misc{harris2013add,
  title={The {Add Health} study: Design and accomplishments},
  author={Harris, Kathleen Mullan},
  howpublished={Carolina Population Center, University of North Carolina at Chapel Hill},
  year={2013}
}

@article{carrell2013natural,
  author = {Carrell, Scott E. and Sacerdote, Bruce I. and West, James E.},
  title = {From Natural Variation to Optimal Policy? {The} Importance of Endogenous Peer Group Formation},
  journal = {Econometrica},
  volume = {81},
  number = {3},
  pages = {855--882},
  year = {2013}
}

@article{karimi2018homophily,
  title={Homophily influences ranking of minorities in social networks},
  author={Karimi, Fariba and G{\'e}nois, Mathieu and Wagner, Claudia and Singer, Philipp and Strohmaier, Markus},
  journal={Scientific Reports},
  volume={8},
  number={1},
  pages={11077},
  year={2018},
  doi={10.1038/s41598-018-29405-7}
}

@article{simon2012collecting,
  title={Collecting ethnic statistics in {Europe}: a review},
  author={Simon, Patrick},
  journal={Ethnic and Racial Studies},
  volume={35},
  number={8},
  pages={1366--1391},
  year={2012},
  doi={10.1080/01419870.2011.607507}
}

@article{smith2014ethnic,
  title={Ethnic ingroup friendships in schools: Testing the by-product hypothesis in {England}, {Germany}, the {Netherlands} and {Sweden}},
  author={Smith, Sanne and Maas, Ineke and {van Tubergen}, Frank},
  journal={Social Networks},
  volume={39},
  pages={33--45},
  year={2014},
  doi={10.1016/j.socnet.2014.04.003}
}

@article{leszczensky2019drives,
  title={What drives ethnic homophily? {A} relational approach on how ethnic identification moderates preferences for same-ethnic friends},
  author={Leszczensky, Lars and Pink, Sebastian},
  journal={American Sociological Review},
  volume={84},
  number={3},
  pages={394--419},
  year={2019},
  doi={10.1177/0003122419846849}
}

@misc{eu2016gdpr,
  author={{European Parliament} and {Council of the European Union}},
  title={Regulation ({EU}) 2016/679 of the {European} {Parliament} and of the {Council} of 27 {April} 2016 on the protection of natural persons with regard to the processing of personal data and on the free movement of such data, and repealing {Directive} 95/46/{EC} ({General} {Data} {Protection} {Regulation})},
  howpublished={Official Journal of the European Union, L 119, 4 May 2016, pp.~1--88},
  year={2016}
}

@article{smith2016ethnic,
  title={Ethnic composition and friendship segregation: Differential effects for adolescent natives and immigrants},
  author={Smith, Sanne and Van Tubergen, Frank and Maas, Ineke and McFarland, Daniel A},
  journal={American Journal of Sociology},
  volume={121},
  number={4},
  pages={1223--1272},
  year={2016},
  publisher={University of Chicago Press, Chicago, IL}
}

@article{kruse2019more,
  title={More than a sorting machine: Ethnic boundary making in a stratified school system},
  author={Kruse, Hanno and Kroneberg, Clemens},
  journal={American Journal of Sociology},
  volume={125},
  number={2},
  pages={431--484},
  year={2019},
  publisher={The University of Chicago Press, Chicago, IL}
}

@article{vit2026rigid,
  title={Rigid boundaries, selective salience: How classroom gender composition shapes adolescent friendships},
  author={Vit, Eszter and Raabe, Isabel J},
  journal={Social Networks},
  volume={86},
  pages={401--411},
  year={2026},
  publisher={Elsevier}
}

\clearpage

\setcounter{page}{1}
\renewcommand{\thepage}{S\arabic{page}}
\setcounter{section}{0}\setcounter{subsection}{0}
\setcounter{equation}{0}\setcounter{figure}{0}\setcounter{table}{0}
\renewcommand{\thesection}{S\arabic{section}}
\renewcommand{\thesubsection}{\thesection.\arabic{subsection}}
\renewcommand{\theequation}{S\arabic{equation}}
\renewcommand{\thefigure}{S\arabic{figure}}
\renewcommand{\thetable}{S\arabic{table}}
\renewcommand{\theHsection}{S\arabic{section}}
\renewcommand{\theHsubsection}{S\arabic{section}.\arabic{subsection}}
\renewcommand{\theHequation}{S\arabic{equation}}
\renewcommand{\theHfigure}{S\arabic{figure}}
\renewcommand{\theHtable}{S\arabic{table}}

\begin{center}
{\Large\bfseries SI Appendix}\\[6pt]
{\large Explaining intersectional segregation in multi-attribute social networks}\\[4pt]
Clara Eminente, Elisa Omodei, Michele Starnini
\end{center}

\section{The model}\label{S-si:model}
This section gives the definitions and properties of the multi-attribute model (MAM) that the main text states in words: the population matrix and its parametrization, the preference matrices, the imbalances of attribute~$B$ within the two groups of attribute~$A$, the densities of ties, and the degree imbalance. Section~\ref{S-si:maci} uses them to calculate the Coleman index.

\subsection{Population}\label{S-si:population}
Each of the $N$ individuals carries two binary attributes, $A\in\{A_0,A_1\}$ and $B\in\{B_0,B_1\}$. For instance, $A$ can be race, with a majority and a minority group (P.~Starr, \emph{Du Bois Review} 20:1--20, 2023), and $B$ gender. The composition of the population is described by the population matrix~$F$, whose entry $F_{a,b}$ is the fraction of individuals of type $(A_a,B_b)$:
\begin{equation}\label{S-eq:Fmatrix}
F=\begin{pmatrix} F_{A_0,B_0} & F_{A_0,B_1}\\ F_{A_1,B_0} & F_{A_1,B_1}\end{pmatrix},\qquad \sum_{a,b}F_{a,b}=1.
\end{equation}
The two marginals
\begin{equation}\label{S-eq:marginals}
\fA=F_{A_0,B_0}+F_{A_0,B_1},\qquad \fB=F_{A_0,B_0}+F_{A_1,B_0}
\end{equation}
are the shares of groups $A_0$ and~$B_0$, with $f_{A_1}=1-\fA$ and $f_{B_1}=1-\fB$. The four entries of~$F$ are determined by three parameters: the two marginals and the consolidation~$k$, the fraction of the members of~$A_0$ who belong to~$B_0$, as parametrized by \citet{martin2025intersectional} (Eq.~(\ref{eq:Fparam}) of the main text). All four entries must be non-negative:
\begin{equation}\label{S-eq:feasibility}
F_{a,b}\geq0\quad\Longleftrightarrow\quad k\,\fA\leq\fB\ \text{ and }\ (1-k)\,\fA\leq1-\fB .
\end{equation}
Under the minority convention $\fA,\fB\leq1/2$ of \citet{martin2025intersectional}, which we adopt, the second condition always holds, and the feasible range of~$k$ is
\begin{equation}\label{S-eq:krange}
0\leq k\leq\min\bigl(1,\fB/\fA\bigr).
\end{equation}
When $k=\fB$, the two attributes are statistically independent: $F_{a,b}=f_{A_a}f_{B_b}$. At the two ends of the feasible range, $k=0$ and $k=\min(1,\fB/\fA)$, they are as anti-correlated and as correlated as the marginals allow. The strength of consolidation is therefore measured by the distance of~$k$ from~$\fB$, in either direction.

\subsection{Preferences}\label{S-si:preferences}
In the model of \citet{martin2026simple}, the tie between individuals $i$ and~$j$ is an independent Bernoulli trial whose probability of success is proportional to $H^A_{A(i),A(j)}\,H^B_{B(i),B(j)}$ (Eq.~(\ref{eq:link_prob}) of the main text), where $H^A$ and~$H^B$ are $2\times2$ preference matrices with free entries; we call this unconstrained version the uMAM. The MAM uses symmetric, row-normalized matrices,
\begin{equation}\label{S-eq:Hmatrices}
H^A=\begin{pmatrix} h_A & 1-h_A\\ 1-h_A & h_A\end{pmatrix},\qquad
H^B=\begin{pmatrix} h_B & 1-h_B\\ 1-h_B & h_B\end{pmatrix},\qquad h_A,h_B\in(0,1),
\end{equation}
and writes them in terms of the centered preferences
\begin{equation}\label{S-eq:alpha_def}
\alpha=2h_A-1,\qquad \beta=2h_B-1,\qquad \alpha,\beta\in(-1,1),
\end{equation}
which measure homophily relative to the neutral value $h_A=h_B=1/2$. Thus $H^A_{a,a}=(1+\alpha)/2$ and $H^A_{a,\bar a}=(1-\alpha)/2$, where $\bar a$ denotes the other group, and similarly for~$B$.

The main text interprets the product form as the acceptance stage of a meet-and-accept process, in which the match is judged once on each attribute. If instead each of the two partners judged the match on each attribute, and both had to accept, the probability of a tie would be proportional to $(H^A_{A(i),A(j)})^2\,(H^B_{B(i),B(j)})^2$. This kernel has the same multiplicative structure: up to an overall scale, it is a MAM with preferences $h'=h^2/[h^2+(1-h)^2]$ in place of~$h$ for each attribute. The multiplicative form thus describes any matching process in which the two attributes are evaluated independently, and it is a natural baseline against which preferences for specific combinations of the two attributes can be tested.

\subsection{Imbalances of attribute $B$ within the $A$-groups}\label{S-si:imbalances}
For each $A$-group, the $B$-imbalance is
\begin{align}
\delta_0 &= F_{A_0,B_0}-F_{A_0,B_1}=(2k-1)\,\fA, \label{S-eq:delta0}\\
\delta_1 &= F_{A_1,B_0}-F_{A_1,B_1}=(2\fB-1)-(2k-1)\,\fA. \label{S-eq:delta1}
\end{align}
It measures how unevenly attribute~$B$ is distributed within the group: $\delta_a=0$ when group~$A_a$ has as many members in~$B_0$ as in~$B_1$. The two imbalances add up to the $B$-imbalance of the population,
\begin{equation}\label{S-eq:delta_sum}
\delta_0+\delta_1=2\fB-1 .
\end{equation}
The cross-imbalance
\begin{equation}\label{S-eq:Delta}
\Delta=\delta_0-\delta_1=2\fA\,(2k-1)-(2\fB-1)
\end{equation}
measures the difference between the $B$-compositions of the two $A$-groups. When $\Delta=0$, each $A$-group accounts for exactly half of the $B$-imbalance of the population ($\delta_0=\delta_1=(2\fB-1)/2$). This does not mean that the two groups have the same internal composition, since their sizes differ; it means that their absolute contributions to the excess of one $B$-group are equal. When $\Delta\neq0$, one $A$-group drives the $B$-imbalance of the population disproportionately. At independence, $\Delta=(2\fB-1)(2\fA-1)$, which vanishes only if $\fA=1/2$ or $\fB=1/2$.

\subsection{Densities of ties}\label{S-si:densities}
With Eqs.~\eqref{S-eq:delta0}--\eqref{S-eq:delta1}, the population matrix reads $F_{a,b}=(f_{A_a}\pm\delta_a)/2$, with the plus sign for $b=0$ and the minus sign for $b=1$. The within-group density of group~$A_a$, the (unnormalized) probability that a randomly chosen ordered pair of individuals belongs to~$A_a$ and is connected, is
\begin{equation}\label{S-eq:Wa}
W_a=\sum_{b,b'}F_{a,b}\,F_{a,b'}\,H^A_{a,a}\,H^B_{b,b'}=\frac{1+\alpha}{2}\Bigl[\frac{f_{A_a}^2}{2}+\frac{\beta}{2}\,\delta_a^2\Bigr],
\end{equation}
and the density of ties within $A$-groups is
\begin{equation}\label{S-eq:W}
W=W_0+W_1=\frac{1+\alpha}{2}\Bigl[\frac{\fA^2+f_{A_1}^2}{2}+\frac{\beta}{2}\,(\delta_0^2+\delta_1^2)\Bigr].
\end{equation}
The density of all the ties sent by members of~$A_a$, to either group, adds the ties to the other group:
\begin{equation}\label{S-eq:Sigma}
\Sigma_a=W_a+\frac{1-\alpha}{2}\Bigl[\frac{f_{A_a}\,f_{A_{\bar a}}}{2}+\frac{\beta}{2}\,\delta_a\,\delta_{\bar a}\Bigr].
\end{equation}
The network density is therefore
\begin{equation}\label{S-eq:Z}
Z=\Sigma_0+\Sigma_1=\frac{1+\alpha}{2}\Bigl[\frac{\fA^2+f_{A_1}^2}{2}+\frac{\beta}{2}\,(\delta_0^2+\delta_1^2)\Bigr]+\frac{1-\alpha}{2}\Bigl[\fA f_{A_1}+\beta\,\delta_0\,\delta_1\Bigr],
\end{equation}
and, with Eqs.~\eqref{S-eq:delta_sum}--\eqref{S-eq:Delta}, it takes the symmetric form
\begin{equation}\label{S-eq:Z_symmetric}
Z=\frac{1}{4}\Bigl[1+\alpha\,(2\fA-1)^2+\beta\,(2\fB-1)^2+\alpha\beta\,\Delta^2\Bigr].
\end{equation}

\begin{figure}[tbp]
\centering
\includegraphics[width=0.35\textwidth]{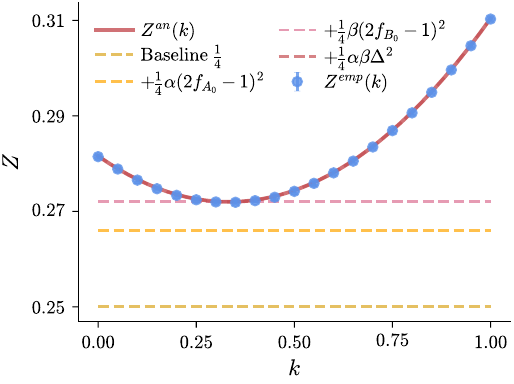}
\caption{\textbf{The network density depends on the consolidation only through the intersectional spillover.} Network density~$Z$ against the consolidation~$k$. Blue symbols, $Z^{emp}$ in the legend: density measured on simulated networks, mean and 95\% confidence interval (often smaller than the symbols). Red line, $Z^{an}$: Eq.~\ref{S-eq:Z_symmetric}. Colored curves: cumulative contributions of the base density $1/4$ (``Baseline'' in the legend), homophily over~$A$, homophily over~$B$, and the intersectional spillover $\alpha\beta\Delta^2/4$. Parameters: $\fA=0.30$, $\fB=0.40$, $\alpha=0.40$, $\beta=0.60$; $N=2{,}000$, 30 networks per point (Materials and Methods, ``Simulations'').}
\label{S-fig:Z}
\end{figure}

The density has four parts. The base density $1/4$ is its value without preferences. The second and third terms are the contributions of homophily over~$A$ and over~$B$; each is a centered preference times the squared size imbalance of its attribute, $(2\fA-1)^2$ or $(2\fB-1)^2$, which measures how far the attribute is from having two groups of equal size. These three terms do not depend on~$k$. The fourth term, $\alpha\beta\Delta^2/4$, is the intersectional spillover. It is the only part of~$Z$ that depends on~$k$; it raises the density when the two preferences have the same sign and lowers it when they have opposite signs, provided that $\Delta\neq0$. When both preferences are homophilic, it reflects the emergence of super-groups: individuals form ties preferentially with others who share their whole type $(A_a,B_b)$. When $\Delta=0$, the spillover vanishes and the density is the sum of two independent homophily effects. Fig.~\ref{S-fig:Z} confirms Eq.~\ref{S-eq:Z_symmetric} on simulated networks.

\subsection{Degree shares and degree imbalance}\label{S-si:gamma}
The \emph{degree share} of group~$A_a$,
\begin{equation}\label{S-eq:phi}
\phi_a=\Sigma_a/Z,
\end{equation}
is the fraction of all ties that are sent by members of~$A_a$; $\phi_0+\phi_1=1$. The Coleman baseline depends on the degree shares (section~\ref{S-si:baseline}). We measure the difference between degree shares and population shares by the \emph{degree imbalance}
\begin{equation}\label{S-eq:gamma}
\gamma=\phi_0-\fA ,
\end{equation}
so that $\phi_1-f_{A_1}=-\gamma$. If $\bar d_0$, $\bar d_1$ and~$\bar d$ are the average numbers of ties of a member of~$A_0$, of a member of~$A_1$, and over the whole network, then $\phi_0=\fA\,\bar d_0/\bar d$ and
\begin{equation}\label{S-eq:gamma_degree}
\gamma=\fA\,(1-\fA)\,\frac{\bar d_0-\bar d_1}{\bar d},
\end{equation}
so that $\gamma>0$ exactly when members of~$A_0$ have, on average, more ties than members of~$A_1$. The degree imbalance is thus a simple, mean-based measure of degree inequality. \citet{martin2025intersectional} use a measure based on the full distributions of the number of ties, which cannot be obtained from~$\gamma$.

From Eqs.~\eqref{S-eq:phi}--\eqref{S-eq:gamma}, $\gamma\,Z=f_{A_1}\Sigma_0-\fA\,\Sigma_1$, and $\gamma$ depends on all the parameters. Switching off one preference at a time isolates two sources of degree imbalance. The first is homophily over~$A$ combined with groups of different size. At $\beta=0$,
\begin{equation}\label{S-eq:gamma_noB}
\gamma\big|_{\beta=0}=\frac{\alpha\,\fA\,(1-\fA)\,(2\fA-1)}{2Z},\qquad Z\big|_{\beta=0}=\frac{1}{4}\Bigl[1+\alpha\,(2\fA-1)^2\Bigr],
\end{equation}
which vanishes for $\alpha=0$ or $\fA=1/2$. With homophily ($\alpha>0$), the smaller group $A_0$ has fewer partners of its own type, so $\phi_0<\fA$ and $\gamma<0$: the larger group is the better connected one. The second source is homophily over~$B$ combined with correlated attributes. At $\alpha=0$,
\begin{equation}\label{S-eq:gamma_alpha0}
\gamma\big|_{\alpha=0}=\frac{\fA\,\beta\,(2\fB-1)\,(k-\fB)}{2Z},\qquad Z\big|_{\alpha=0}=\frac{1}{4}\Bigl[1+\beta\,(2\fB-1)^2\Bigr],
\end{equation}
which vanishes for $\beta=0$, $\fB=1/2$, or $k=\fB$. Here the two $A$-groups differ only in their $B$-composition. With homophily over~$B$ ($\beta>0$), the members of the larger $B$-group have more ties, and so does the $A$-group that is over-represented in it: the factor $2\fB-1$ identifies the larger $B$-group, the factor $k-\fB$ the $A$-group that is over-represented in~$B_0$, and the sign of~$\gamma$ changes as $k$ crosses~$\fB$.

\section{The Coleman index in the model}\label{S-si:maci}
This section derives the results on the Coleman index stated in the main text (Results, ``The Coleman index of one attribute mixes two preferences'' and ``Correlated attributes can induce segregation'', and Materials and Methods, ``Coleman index in the model''): the value of the Coleman baseline in the model, the closed form of the index, the identity for $\chiA-\alpha$ (Eq.~(\ref{eq:chiA_minus_alpha}) of the main text), and the fact that induced segregation is not a by-product of differences in the number of ties. It also gives the formulas for attribute~$B$ and the numerical validation of the closed form.

\subsection{The Coleman baseline in the model}\label{S-si:baseline}
In the MAM and for large~$N$, the Coleman baseline, Eq.~(\ref{eq:omega_star}) of the main text, equals
\begin{equation}\label{S-eq:wstar_lemma}
w_A^*=\fA\,\phi_0+(1-\fA)\,\phi_1 ,
\end{equation}
where $\phi_g$ is the degree share of group~$A_g$, Eq.~\ref{S-eq:phi}. Each tie leaves a member of~$A_g$ with probability~$\phi_g$ and, under the baseline, reaches a member of the same group with probability~$f_{A_g}$.

To derive Eq.~\ref{S-eq:wstar_lemma}, write the baseline as $w_A^*=\omega_A^*/M$, where $M=\sum_i\eta_i$ is the total number of ties and $\omega_A^*=\sum_g\sum_{i\in G_g}\eta_i\,(n_g-1)/(N-1)$ is the expected number of within-group ties. For large~$N$, $(n_g-1)/(N-1)\to f_{A_g}$, so that $\omega_A^*\to\sum_g M_g\,f_{A_g}$, where $M_g=\sum_{i\in G_g}\eta_i$ is the total out-degree of group~$A_g$. In the MAM, $M_g$ is proportional to~$\Sigma_g$, with the same constant of proportionality for both groups, and therefore $M=M_0+M_1$ is proportional to~$Z$. Hence $w_A^*=\sum_g f_{A_g}\,\Sigma_g/Z=\sum_g f_{A_g}\,\phi_g$. When the expected number of ties does not depend on the type of an individual, $\Sigma_g$ is proportional to~$f_{A_g}$, so that $\phi_g=f_{A_g}$ and the baseline reduces to the random-mixing value of Eq.~\ref{S-eq:wstar0} below.

On the scale of the densities of section~\ref{S-si:densities}, the baseline is
\begin{equation}\label{S-eq:Wstar}
W^*=Z\,w_A^*=\fA\,\Sigma_0+f_{A_1}\,\Sigma_1 ,
\end{equation}
and, since $\Sigma_0+\Sigma_1=Z$, the expected density of cross-group ties under the baseline is $Z-W^*=f_{A_1}\Sigma_0+\fA\,\Sigma_1>0$.

\subsection{The random-mixing value and the degree imbalance}\label{S-si:randommixing}
The \emph{random-mixing value}
\begin{equation}\label{S-eq:wstar0}
w_A^{*,0}=\fA^2+(1-\fA)^2
\end{equation}
is the fraction of within-group ties expected when partners are chosen at random and all individuals have the same number of ties; it is the aggregate counterpart of the ``baseline homophily'' of \citet{currarini2009economic}. The Coleman index does not use it. Substituting $\phi_0=\fA+\gamma$, Eq.~\ref{S-eq:gamma}, into Eq.~\ref{S-eq:wstar_lemma} gives the identity
\begin{equation}\label{S-eq:1mwstar_gamma}
w_A^*=w_A^{*,0}+\gamma\,(2\fA-1),\qquad\text{equivalently}\qquad 1-w_A^*=2\fA(1-\fA)-\gamma\,(2\fA-1).
\end{equation}
The Coleman baseline thus differs from the random-mixing value by a term due to the degree imbalance, which vanishes when $\fA=1/2$. Eq.~\ref{S-eq:1mwstar_gamma} is a decomposition: the index is defined with~$w_A^*$, and the identity only shows how $w_A^*$ differs from the random-mixing value.

\paragraph{A single attribute.} With no preference over~$B$ ($\beta=0$) and homophily over~$A$ ($\alpha>0$), the index equals~$\alpha$ (section~\ref{S-si:chiA_minus_alpha}). Computed with $w_A^{*,0}$ in place of~$w_A^*$, it would instead be
\begin{equation}\label{S-eq:chi0_single}
\chi_A^{0}=\alpha\,\frac{1+(2\fA-1)^2}{1+\alpha\,(2\fA-1)^2},
\end{equation}
which exceeds~$\alpha$ whenever $\fA\neq1/2$. For $\fA=0.2$ and $\alpha=0.6$, $\chi_A^0=0.6\times1.36/1.216\approx0.67$. To derive Eq.~\ref{S-eq:chi0_single}, note that at $\beta=0$ Eqs.~\eqref{S-eq:W} and~\eqref{S-eq:gamma_noB} give $W/Z=(1+\alpha)(1+u)/[2(1+\alpha u)]$ with $u=(2\fA-1)^2$, and $w_A^{*,0}=(1+u)/2$; then $\chi_A^0=(W/Z-w_A^{*,0})/(1-w_A^{*,0})$.

\paragraph{Effect of the degree imbalance on the index.} Let $g=\gamma\,(2\fA-1)/[2\fA(1-\fA)]$ be the relative reduction, due to the degree imbalance, of the expected fraction of cross-group ties. When both $\chiA$ and $\chi_A^0$ are in the homophilic regime, Eq.~\ref{S-eq:1mwstar_gamma} gives
\begin{equation}\label{S-eq:chi_gamma_identity}
1-\chiA=\frac{1-\chi_A^{0}}{1-g},
\end{equation}
because $1-\chiA=(1-w_A)/(1-w_A^*)$ and $1-\chi_A^0=(1-w_A)/(1-w_A^{*,0})$. If the larger group is also the better connected one ($g>0$), fewer cross-group ties are expected by chance, so the same observed mixing corresponds to less segregation than $\chi_A^{0}$ indicates; if the smaller group is the better connected one ($g<0$), the opposite holds. The identity follows from the definition of the index alone, and holds for any network.

\subsection{Closed form}\label{S-si:closedform}
In the model, the expected fraction of within-group ties is $W/Z$, and the index, Eq.~(\ref{eq:CI}) of the main text, reads $\chiA=(W-W^*)/(Z-W^*)$ in the homophilic regime ($W\geq W^*$) and $\chiA=(W-W^*)/W^*$ in the heterophilic regime ($W<W^*$). Computing $W-W^*$ from Eqs.~\eqref{S-eq:W} and~\eqref{S-eq:Wstar}, and simplifying, the numerator is the sum of a direct and an indirect term:
\begin{equation}\label{S-eq:numA_general}
W-W^*=\underbrace{\frac{\alpha}{2}\,\fA\,(1-\fA)}_{\text{direct term}}+\underbrace{\frac{\beta}{2}\,Q_A}_{\text{indirect term}},
\end{equation}
where $Q_A=\tfrac{1+\alpha}{2}\bigl[(1-\fA)\,\delta_0^2+\fA\,\delta_1^2\bigr]-\tfrac{1-\alpha}{2}\,\delta_0\,\delta_1$. The denominator of the homophilic regime has the same structure:
\begin{equation}\label{S-eq:denA_general}
Z-W^*=\frac{1}{2}\,\fA\,(1-\fA)+\frac{\beta}{2}\,\Qt_A,
\end{equation}
where $\Qt_A=\tfrac{1+\alpha}{2}\bigl[(1-\fA)\,\delta_0^2+\fA\,\delta_1^2\bigr]+\tfrac{1-\alpha}{2}\,\delta_0\,\delta_1$, so that $\Qt_A-Q_A=(1-\alpha)\,\delta_0\,\delta_1$. In the homophilic regime the index is therefore
\begin{equation}\label{S-eq:chiA_closed}
\chiA=\frac{\tfrac{\alpha}{2}\,\fA(1-\fA)+\tfrac{\beta}{2}\,Q_A}{\tfrac{1}{2}\,\fA(1-\fA)+\tfrac{\beta}{2}\,\Qt_A}\qquad(W\geq W^*).
\end{equation}
Multiplying numerator and denominator by~2 gives Eqs.~(\ref{eq:chiA_closedform}) and~(\ref{eq:QA}) of the main text. Each of the two terms of the numerator is gated by one preference: setting $\beta=0$ removes the indirect term, and setting $\alpha=0$ removes only the direct term. The direct term is a single-attribute effect; the indirect term is where the preference over~$B$ enters the index of~$A$. At $\beta=0$ the denominator is $\tfrac12\fA(1-\fA)$, and the index equals~$\alpha$.

In the heterophilic regime, the denominator is $W^*$, which by Eqs.~\eqref{S-eq:Sigma} and~\eqref{S-eq:Wstar} reads
\begin{equation}\label{S-eq:Wstar_expanded}
W^*=\frac{1+\alpha}{2}\Bigl[\frac{\fA^3+f_{A_1}^3}{2}+\frac{\beta}{2}\,\bigl(\fA\,\delta_0^2+f_{A_1}\,\delta_1^2\bigr)\Bigr]+\frac{1-\alpha}{2}\Bigl[\frac{\fA f_{A_1}}{2}+\frac{\beta}{2}\,\delta_0\,\delta_1\Bigr],
\end{equation}
so that
\begin{equation}\label{S-eq:chiA_hetero}
\chiA=\frac{\tfrac{\alpha}{2}\,\fA(1-\fA)+\tfrac{\beta}{2}\,Q_A}{W^*}\qquad(W<W^*).
\end{equation}
The two regimes are related by a scalar rescaling,
\begin{equation}\label{S-eq:rescaling_identity}
\chi_A^{\rm hetero}=\chi_A^{\rm homo}\,\frac{Z-W^*}{W^*},
\end{equation}
where $\chi_A^{\rm homo}$ denotes the expression~\eqref{S-eq:chiA_closed}; the homophilic formula therefore determines both regimes.

\subsection{When the index equals the preference}\label{S-si:chiA_minus_alpha}
In the homophilic regime, the index differs from the preference~$\alpha$ by
\begin{equation}\label{S-eq:chiA_minus_alpha_SI}
\chiA-\alpha=\frac{\beta\,(1-\alpha^2)\,\fA\,(k-\fB)\,\Delta}{2\,(Z-W^*)},
\end{equation}
which is Eq.~(\ref{eq:chiA_minus_alpha}) of the main text, since $2(Z-W^*)=\fA(1-\fA)+\beta\,\Qt_A$ by Eq.~\ref{S-eq:denA_general}. Since $Z-W^*>0$ and $|\alpha|<1$, the index differs from~$\alpha$ unless $\beta=0$, $k=\fB$ (independent attributes) or $\Delta=0$.

To derive Eq.~\ref{S-eq:chiA_minus_alpha_SI}, subtract $\alpha$ times the denominator~\eqref{S-eq:denA_general} from the numerator~\eqref{S-eq:numA_general}:
\[
(W-W^*)-\alpha\,(Z-W^*)=\frac{\beta}{2}\,\bigl(Q_A-\alpha\,\Qt_A\bigr)=\frac{\beta}{4}\,(1-\alpha^2)\,\bigl[(1-\fA)\,\delta_0^2+\fA\,\delta_1^2-\delta_0\,\delta_1\bigr].
\]
The bracket factorizes as $(\delta_0-\delta_1)\,\bigl[(1-\fA)\,\delta_0-\fA\,\delta_1\bigr]$, and Eqs.~\eqref{S-eq:delta0}--\eqref{S-eq:delta1} give $(1-\fA)\,\delta_0-\fA\,\delta_1=2\fA\,(k-\fB)$, so that the bracket equals $2\fA\,(k-\fB)\,\Delta$. Dividing by $Z-W^*$ gives Eq.~\ref{S-eq:chiA_minus_alpha_SI}. In the second case, $k=\fB$, one has $\delta_0=\fA(2\fB-1)$ and $\delta_1=(1-\fA)(2\fB-1)$, so that $Q_A=\alpha\,(2\fB-1)^2\fA(1-\fA)$ and $\Qt_A=Q_A/\alpha$. In the third case, $\Delta=0$, one has $\delta_0=\delta_1$, $Q_A=\alpha\,\delta_0^2$ and $\Qt_A=\delta_0^2$. In both cases, the indirect terms of numerator and denominator stand in the ratio~$\alpha$, and the index equals~$\alpha$; the indirect term of the numerator does not vanish, but it is offset by that of the denominator. At $\alpha=0$, Eq.~\ref{S-eq:chiA_minus_alpha_SI} gives the numerator $W-W^*=\tfrac{\beta}{2}\,\fA\,(k-\fB)\,\Delta$, which is used in section~\ref{S-si:alpha0}.

\subsection{Induced segregation with equal numbers of ties}\label{S-si:equalties}
With no preference over~$A$ and two $B$-groups of the same size ($\alpha=0$, $\fB=1/2$), every individual has the same expected number of ties, yet the numerator of the index of~$A$ is
\begin{equation}\label{S-eq:num_equalties}
W-W^*=\frac{\beta}{2}\,\fA^2\,(2k-1)^2 ,
\end{equation}
which is positive with homophily over~$B$ ($\beta>0$) whenever the attributes are correlated ($k\neq1/2$). Induced segregation is therefore not a by-product of differences in the number of ties.

At $\alpha=0$ the matrix $H^A$ has all its entries equal to $1/2$, so the expected number of ties of an individual in group~$B_b$ is proportional to $\tfrac12\sum_{b'}f_{B_{b'}}H^B_{b,b'}$, whatever its $A$-group. At $\fB=1/2$ this sum equals $1/2$ for both $B$-groups, so that every individual has the same expected number of ties, $\phi_g=f_{A_g}$, $\gamma=0$, and the Coleman baseline equals the random-mixing value~\eqref{S-eq:wstar0}. At $\fB=1/2$, Eq.~\ref{S-eq:Delta} gives $\Delta=2\fA(2k-1)$, and the numerator at $\alpha=0$ of section~\ref{S-si:chiA_minus_alpha} becomes Eq.~\ref{S-eq:num_equalties}. The degree imbalance~\eqref{S-eq:gamma_alpha0} requires two $B$-groups of different size; induced segregation does not, because it is due to the different $B$-composition of the two $A$-groups. This is the case of Fig.~\ref{fig:induced}A--B of the main text ($\fA=0.45$, $\fB=0.5$, $k=0.92$, $\beta=0.92$).

\subsection{Attribute $\mathbf{B}$}\label{S-si:attrB}
Swapping the roles of the two attributes, let $\delta'_0=F_{A_0,B_0}-F_{A_1,B_0}=2k\fA-\fB$ and $\delta'_1=F_{A_0,B_1}-F_{A_1,B_1}=2(1-k)\fA-(1-\fB)$ be the $A$-imbalances within the two $B$-groups, with $\delta'_0+\delta'_1=2\fA-1$. The numerator and the homophilic denominator of the index of~$B$ are
\begin{align}
W_B-W_B^* &= \frac{\beta}{2}\,\fB\,(1-\fB)+\frac{\alpha}{2}\,Q_B, \label{S-eq:numB_general}\\
Z-W_B^* &= \frac{1}{2}\,\fB\,(1-\fB)+\frac{\alpha}{2}\,\Qt_B, \label{S-eq:denB_general}
\end{align}
with
\begin{align}
Q_B &= \tfrac{1+\beta}{2}\Bigl[(1-\fB)\,{\delta'_0}^2+\fB\,{\delta'_1}^2\Bigr]-\tfrac{1-\beta}{2}\,\delta'_0\,\delta'_1, \label{S-eq:QB}\\
\Qt_B &= \tfrac{1+\beta}{2}\Bigl[(1-\fB)\,{\delta'_0}^2+\fB\,{\delta'_1}^2\Bigr]+\tfrac{1-\beta}{2}\,\delta'_0\,\delta'_1. \label{S-eq:QtB}
\end{align}
The indirect term is proportional to~$\beta$ in the index of~$A$, and to~$\alpha$ in the index of~$B$.

\subsection{Numerical validation}\label{S-si:validation}
\begin{figure}[tbp]
\centering
\includegraphics[width=0.35\textwidth]{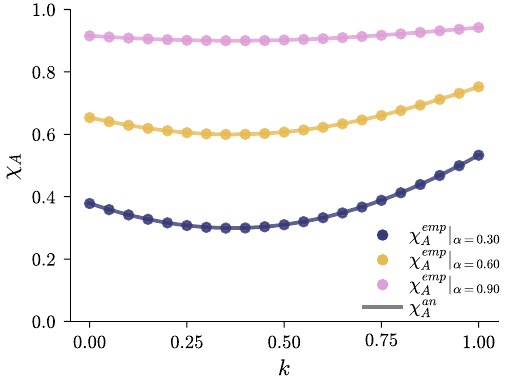}
\caption{\textbf{The closed form agrees with simulations.} Coleman index of~$A$ against the consolidation~$k$ for three values of~$\alpha$, measured on networks simulated from the MAM (symbols, $\chi_A^{emp}$ in the legend; 30 networks of $N=2{,}000$ individuals per point) and calculated with Eq.~(\ref{eq:chiA_closedform}) of the main text (lines, $\chi_A^{an}$), for $\fA=0.30$, $\fB=0.40$ and $\beta=0.40$. The index equals~$\alpha$ at independence ($k=\fB$) and departs from it as $k$ moves away.}
\label{S-fig:validation}
\end{figure}

Fig.~\ref{S-fig:validation} validates the closed form~\eqref{S-eq:chiA_closed} for three values of~$\alpha$ (0.3, 0.6 and 0.9) at $\beta=0.4$, $\fA=0.3$ and $\fB=0.4$. The index measured on simulated networks (30 networks of $N=2{,}000$ individuals per point; Materials and Methods, ``Simulations'') falls on the calculated curves over the whole feasible range of the consolidation, $0\leq k\leq1$ for these shares. This confirms the full closed form, including the terms in $Q_A$ and~$\Qt_A$ that depend on both preferences.

\section{One neutral preference}\label{S-si:alpha0}
This section derives the results of the main text for the case without preference over attribute~$A$ ($\alpha=0$; Results, ``With one neutral preference, only induced segregation can be removed''): the two indices at $\alpha=0$, the splits that remove segregation on~$A$, the existence condition of the equal-size decorrelating split (ESDS), and the planner's optimum. It ends with a numerical comparison of the ESDS and a random split. The tests on the data are in section~\ref{S-si:splittests}.

\subsection{The index of $A$ at $\alpha=0$}\label{S-si:numA_alpha0}
Without preference over~$A$, the direct term of the numerator~\eqref{S-eq:numA_general} vanishes, and section~\ref{S-si:chiA_minus_alpha} gives
\begin{equation}\label{S-eq:numA_half_factored}
W-W^*\big|_{\alpha=0}=\frac{\beta}{2}\,\fA\,(k-\fB)\,\Delta .
\end{equation}
All the segregation on~$A$ is then induced by the preference over~$B$. For $\beta\neq0$, it vanishes only when the attributes are independent ($k=\fB$) or when $\Delta=0$. In the homophilic regime, dividing by $Z-W^*=\tfrac12[\fA(1-\fA)+\beta\Qt_A]$ gives the index of~$A$ at $\alpha=0$ quoted in the main text, Eq.~(\ref{eq:chiA_minus_alpha}) at $\alpha=0$. In the heterophilic regime the denominator is~$W^*$, but the numerator is the same, so the index has the same sign and the same zeros.

\subsection{The index of $B$ at $\alpha=0$}\label{S-si:chiB_alpha0}
At $\alpha=0$, the Coleman index of~$B$ equals~$\beta$ for every composition $(\fA,\fB,k)$ when the preference over~$B$ is homophilic ($\beta>0$). When it is heterophilic ($\beta<0$), the index depends on the composition only through~$\fB$, and it equals~$\beta$ only when $\fB=1/2$.

To derive these statements, set $\alpha=0$ in the formulas for attribute~$B$, Eqs.~\eqref{S-eq:numB_general}--\eqref{S-eq:denB_general}. Both indirect terms vanish, since their coefficient is $\alpha/2$, and
\[
W_B-W_B^*=\frac{\beta}{2}\,\fB\,(1-\fB),\qquad Z-W_B^*=\frac{1}{2}\,\fB\,(1-\fB).
\]
For $\beta>0$ the homophilic regime applies, and the index is the ratio of the two, $\chiB=\beta$. For $\beta<0$ the heterophilic regime applies, and the denominator is $W_B^*$ itself. At $\alpha=0$ the tie probability is $\tfrac12H^B_{b,b'}$, so the density of ties sent by the members of~$B_b$ is $\Sigma'_b=\tfrac12\,T_b\,M_b$, with $T_0=\fB$, $T_1=1-\fB$ and $M_b=\tfrac{1+\beta}{2}\,T_b+\tfrac{1-\beta}{2}\,T_{1-b}$. Then $W_B^*=T_0\Sigma'_0+T_1\Sigma'_1=\tfrac12\,(T_0^2M_0+T_1^2M_1)$, which depends on~$\fB$ and~$\beta$ but not on~$\fA$ or~$k$. At $\fB=1/2$, $T_0=T_1=M_0=M_1=1/2$, so $W_B^*=1/8$, $W_B-W_B^*=\beta/8$, and $\chiB=(W_B-W_B^*)/W_B^*=\beta$, as in the homophilic regime. When $\fB\neq1/2$, $W_B^*$ varies with~$\fB$, and so does the index.

The intuition is that at $\alpha=0$ attribute~$A$ contributes the same factor to every tie, so from the standpoint of~$B$ the model is a single-attribute model with preference~$\beta$, in which the Coleman baseline accounts exactly for the different numbers of ties of the two $B$-groups (section~\ref{S-si:randommixing}).

\subsection{Two-class splits}\label{S-si:splits}
This subsection states two properties of splits that the rest of this section uses. Splits, the conservation of the composition of the school and the class averages $\langle\chiA\rangle$ and~$\langle\chiB\rangle$ are defined in Materials and Methods, ``Two-class splits''; we write $\lambda=N^{c_1}/N$ for the relative size of class~$c_1$. Within each class, ties follow the MAM with the same preferences $\alpha$ and~$\beta$. Since the Coleman index in the MAM depends on the composition of a class and not on its size, two classes with the composition of the school have the indices of the undivided school.

At $\alpha=0$ and $\beta>0$, section~\ref{S-si:chiB_alpha0} gives $\chiB^{c_j}=\beta$ in each class, whatever its composition, so $\langle\chiB\rangle=\beta$ for every split: the planner cannot change segregation on~$B$ through the composition of the classes (the policy impossibility of the main text).

\subsection{Splits that remove segregation on $A$}\label{S-si:esds}
By Eq.~\ref{S-eq:numA_half_factored}, the index of~$A$ is zero in any class in which the two attributes are independent, $k^{c_j}=\fB^{c_j}$. Two kinds of split achieve this.

\paragraph{The $B$-stratified split.} If the school contains at least one student of each $B$-group, assigning all the students of~$B_0$ to class~$c_1$ and all those of~$B_1$ to class~$c_2$ gives $\chiA^{c_j}=0$ in each class, and hence $\langle\chiA\rangle=0$. Indeed, the conservation constraint holds by construction. In class~$c_1$ every student belongs to~$B_0$, so $\fB^{c_1}=1$, and every member of~$A_0$ in it also belongs to~$B_0$, so $k^{c_1}=1$; hence $k^{c_1}-\fB^{c_1}=0$ and the numerator~\eqref{S-eq:numA_half_factored} vanishes, whatever the value of~$\Delta^{c_1}$. In class~$c_2$, $\fB^{c_2}=0$ and $k^{c_2}=0$, and the same holds. Each class, however, contains a single $B$-group, so $\fB^{c_j}(1-\fB^{c_j})=0$ and the index of~$B$ is not defined within the classes. One can read this as complete segregation on~$B$: the planner obtains $\langle\chiA\rangle=0$ by giving up all control over segregation on~$B$.

\paragraph{The decorrelating split.} A milder split keeps both $B$-groups in each class and makes the attributes independent within each class, $k^{c_j}=\fB^{c_j}$. It requires a decomposition $F=\lambda F^{c_1}+(1-\lambda)F^{c_2}$ in which each $F^{c_j}$ is the product of its marginals (equivalently, has rank~1). Not every school admits one. With two classes of equal size ($\lambda=1/2$), such a decomposition exists if and only if
\begin{equation}\label{S-eq:rank1_condition}
0\leq k\leq2\fB,\qquad\text{equivalently}\qquad |k-\fB|\leq\fB,
\end{equation}
and the solutions form a one-parameter family. In terms of the normalized consolidation $\kappa=(k-\fB)/\fB\in[-1,1]$ and the relative deviations of the marginals of class~$c_1$,
\[
t=\frac{\fA^{c_1}-\fA}{\fA},\qquad u=\frac{\fB^{c_1}-\fB}{\fB},
\]
all equal-size decorrelating splits are obtained by choosing $t$ with $|\kappa|\leq|t|\leq1$, $t\neq0$, and setting
\begin{equation}\label{S-eq:rank1_param}
u=\kappa/t,\qquad \fA^{c_1}=\fA(1+t),\quad \fB^{c_1}=\fB(1+u),\qquad \fA^{c_2}=\fA(1-t),\quad \fB^{c_2}=\fB(1-u).
\end{equation}
At the boundary $|\kappa|=1$, the only solutions have $|u|=1$, so one class contains a single $B$-group; a decorrelating split that keeps both $B$-groups in each class requires $|\kappa|<1$, that is, $0<k<2\fB$. We call the equal-size decorrelating split the ESDS. When condition~\eqref{S-eq:rank1_condition} fails ($k>2\fB$), the $B$-stratified split still gives $\langle\chiA\rangle=0$.

To derive condition~\eqref{S-eq:rank1_condition}, note that a matrix of rank~1 with nonnegative entries is determined by its marginals: $F^{c_j}_{A_0,B_0}=\fA^{c_j}\fB^{c_j}$, $F^{c_j}_{A_0,B_1}=\fA^{c_j}(1-\fB^{c_j})$, and so on. The decomposition $F=\tfrac12(F^{c_1}+F^{c_2})$ then requires (i) $\tfrac12(\fA^{c_1}+\fA^{c_2})=\fA$, (ii) $\tfrac12(\fB^{c_1}+\fB^{c_2})=\fB$, and (iii) $\tfrac12(\fA^{c_1}\fB^{c_1}+\fA^{c_2}\fB^{c_2})=k\fA$; the other three entries follow. Eliminating $\fA^{c_2}=2\fA-\fA^{c_1}$ and $\fB^{c_2}=2\fB-\fB^{c_1}$ from (iii) gives
\begin{equation}\label{S-eq:rank1_hyperbola}
(\fA^{c_1}-\fA)\,(\fB^{c_1}-\fB)=\fA\,(k-\fB),
\end{equation}
that is, $t\,u=\kappa$. Under the minority convention, the requirement that all marginals lie between 0 and~1 reduces to $(t,u)\in[-1,1]^2$. The hyperbola $tu=\kappa$ meets this square if and only if $|\kappa|\leq1$, which is condition~\eqref{S-eq:rank1_condition}. Conversely, if $|\kappa|\leq1$, any $t$ with $|\kappa|\leq|t|\leq1$ gives $u=\kappa/t\in[-1,1]$, and Eq.~\ref{S-eq:rank1_param} defines two rank-1 matrices whose average is~$F$.

Fig.~\ref{fig:esdsmap} of the main text shows the condition in the $(\fB,k)$ plane, together with the CILS4EU classes used to test it (section~\ref{S-si:splittests}).

\subsection{The planner's optimum}\label{S-si:optimum_alpha0}
At $\alpha=0$ and $\beta>0$, for any school with $0<k<2\fB$, the minimum of the aggregate objective over all splits is $\min J=\beta$. It is attained by any split that brings the index of~$A$ to zero in every class while keeping both $B$-groups in each class, for instance the ESDS. The floor~$\beta$ cannot be lowered.

The derivation is short. By section~\ref{S-si:splits}, $\langle\chiB\rangle=\beta$ for every split, so $|\langle\chiB\rangle|=\beta$ is fixed. The only term left is $|\langle\chiA\rangle|\geq0$, which any decorrelating split brings to zero. For $0<k<2\fB$, section~\ref{S-si:esds} gives an ESDS with both $B$-groups in each class (take $|\kappa|<|t|<1$, so that $|u|<1$), in which $\chiB^{c_j}=\beta$ is defined. The $B$-stratified split also gives $\langle\chiA\rangle=0$, but it leaves $\chiB$ undefined and cannot be used. Hence $\min J=\beta$. In the ESDS each class has $\chiA^{c_j}=0$ and $\chiB^{c_j}=\beta>0$, so $J_{\rm strict}=J=\beta$, and the ESDS is optimal for both objectives.

\subsection{The ESDS and a random split on simulated networks}\label{S-si:esds_numerics}
\begin{figure}[tbp]
\centering
\includegraphics[width=0.65\textwidth]{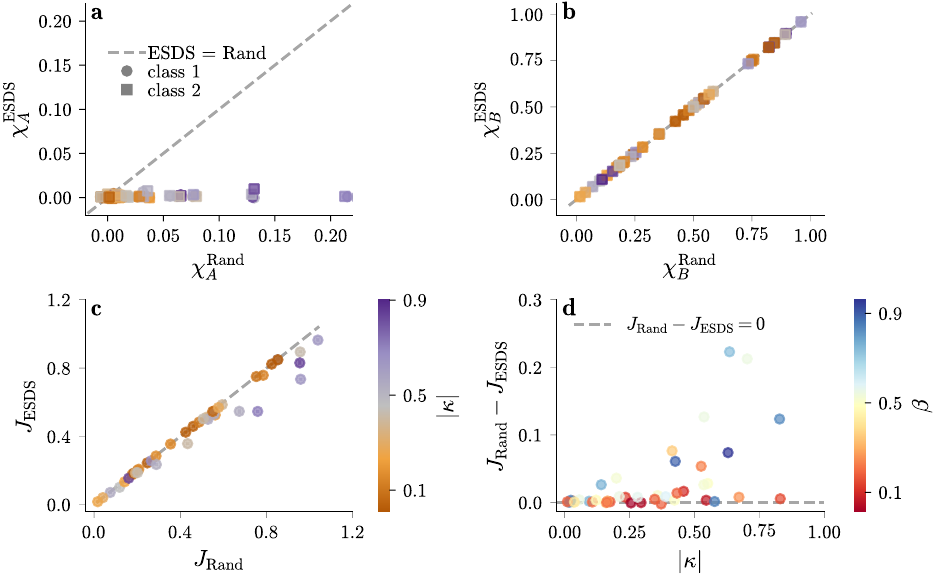}
\caption{\textbf{The advantage of the ESDS over a random split grows with consolidation.} Simulated schools at $\alpha=0$ ($N=1{,}000$; 40 schools; 3 networks per school and 20 random splits per network). (a)~Index of~$A$ in each class (circles: class~1; squares: class~2) after the ESDS (vertical axis) and after a random split (horizontal axis): the ESDS gives zero, the random split does not. (b)~The same for the index of~$B$: both splits give the same value. (c)~Aggregate objective~$J$ after the ESDS against a random split, colored by~$|\kappa|$. (d)~Difference between the objectives of the two splits against~$|\kappa|$, colored by~$\beta$; it grows with~$|\kappa|$.}
\label{S-fig:esds_vs_random}
\end{figure}

A random split, in which every student is assigned to either class at random, gives each class, on average, the composition of the school, so each class keeps $k^{c_j}\approx k$ and $\kappa^{c_j}\approx\kappa$. Induced segregation on~$A$ is then not removed, whereas the ESDS sets $\kappa^{c_j}=0$ by construction. The excess of segregation of the random split over the ESDS should therefore grow with~$|\kappa|$: when the attributes are nearly independent, there is little to decorrelate and the two splits perform alike; when they are strongly consolidated, the ESDS is needed. Fig.~\ref{S-fig:esds_vs_random} confirms this on simulated schools at $\alpha=0$.

\section{Two active preferences in a balanced school}\label{S-si:C12}
This section derives the results of the main text for a school balanced on both attributes, $\fA=\fB=1/2$ (Results, ``With two active preferences, splitting helps only when they conflict'' and ``A school average can hide classes that are more segregated''): the indices as functions of the squared consolidation, the optimal split at independence, the optimal split judged class by class away from independence, the cancellation trap, and the conditions under which both classes improve. Throughout, $s=(2k-1)^2$ is the squared consolidation, the indices are given by Eq.~(\ref{eq:chiAB_C3_s}) of the main text, and splits have two classes of equal size that keep the shares of the school, so that each class is also balanced. The data test is in section~\ref{S-si:splittests}.

\subsection{The indices in a balanced school}\label{S-si:C12_index}
When the two $A$-groups have the same size, the degree shares do not affect the Coleman baseline: Eq.~\ref{S-eq:wstar_lemma} gives $w_A^*=\tfrac12(\phi_0+\phi_1)=\tfrac12$, since $\phi_0+\phi_1=1$ whatever the composition, so $W^*=Z/2$ and $Z-W^*=W^*$. The denominators of the two regimes coincide, and a single formula covers both. At $\fA=\fB=1/2$, Eqs.~\eqref{S-eq:delta0}--\eqref{S-eq:delta1} give $\delta_0=-\delta_1=(2k-1)/2$, so that $Q_A=s/4$ and $\Qt_A=\alpha s/4$. The numerator~\eqref{S-eq:numA_general} is then $(\alpha+\beta s)/8$ and the denominator~\eqref{S-eq:denA_general} is $(1+\alpha\beta s)/8$, which gives $\chiA(s)$ of Eq.~(\ref{eq:chiAB_C3_s}); $\chiB(s)$ follows by exchanging the roles of the attributes.

\paragraph{Properties.} (i)~The two indices are symmetric: $\chiB(\alpha,\beta,s)=\chiA(\beta,\alpha,s)$. (ii)~At independence ($s=0$), $\chiA=\alpha$ and $\chiB=\beta$. (iii)~Under perfect consolidation ($s=1$, i.e.\ $k=0$ or $k=1$), $\chiA=\chiB=(\alpha+\beta)/(1+\alpha\beta)$, which is symmetric in $\alpha$ and~$\beta$. All three follow by substitution.

\paragraph{The pairs of indices lie on a segment.} As $s$ goes from 0 to 1, the pair $(\chiA,\chiB)$ moves along the straight segment from $(\alpha,\beta)$ to the common value of (iii). Indeed,
\[
\frac{d\chiA}{ds}=\frac{\beta\,(1-\alpha^2)}{(1+\alpha\beta s)^2},\qquad
\frac{d\chiB}{ds}=\frac{\alpha\,(1-\beta^2)}{(1+\alpha\beta s)^2},
\]
so the slope $d\chiB/d\chiA=\alpha(1-\beta^2)/[\beta(1-\alpha^2)]$ does not depend on~$s$. Under a split that keeps the shares of the school, each class is also balanced, and the size-weighted averages of the class indices are averages of two points of the segment; since the segment is straight, every attainable pair $(\langle\chiA\rangle,\langle\chiB\rangle)$ lies on it.

\paragraph{Phases.} In the agreement phase ($\alpha\beta>0$), both derivatives have the sign of the indices, so neither index vanishes and both grow in absolute value with~$s$. In the conflict phase ($\alpha\beta<0$), with the labeling $|\alpha|\leq|\beta|$, $\chiA(s)=0$ at $s^*=-\alpha/\beta=|\alpha|/|\beta|\in[0,1]$, where the denominator is $1-\alpha^2>0$, so $\chiA$ changes sign there. The index of~$B$ keeps the sign of~$\beta$ for all $s\in[0,1]$: it equals $\beta$ at $s=0$ and $(\alpha+\beta)/(1+\alpha\beta)$ at $s=1$, and $\alpha+\beta$ has the sign of~$\beta$ because $|\beta|\geq|\alpha|$.

\paragraph{Concavity in the signed consolidation.} In the conflict phase, with $\alpha>0>\beta$, write $v=2k-1\in[-1,1]$ for the signed consolidation of a class, so that $s=v^2$, and $g(v)=\chiA(v^2)$ for its index of~$A$. Then $g$ is strictly concave on $[-1,1]$. By the chain rule, $g''(v)=2\chiA'(v^2)+4v^2\chiA''(v^2)$. The first derivative, $\chiA'(s)=\beta(1-\alpha^2)/(1+\alpha\beta s)^2$, is negative because $\beta<0$, and differentiating once more,
\[
\chiA''(s)=-\frac{2\alpha\beta^2(1-\alpha^2)}{(1+\alpha\beta s)^3}<0 ,
\]
so both terms of $g''$ are negative. Under an equal-size split that keeps the shares, the two classes have signed consolidations $u\pm2\varepsilon$, with $u=2k-1$, so that $\langle\chiA\rangle=\tfrac12[g(u+2\varepsilon)+g(u-2\varepsilon)]$. Since $g'$ is decreasing, $d\langle\chiA\rangle/d\varepsilon=g'(u+2\varepsilon)-g'(u-2\varepsilon)<0$ for $\varepsilon>0$: the average index of~$A$ decreases strictly as the offset grows, and it is never larger than its value $g(u)$ in the undivided school. Sections~\ref{S-si:C12_strict} and~\ref{S-si:C12_trap} use this property.

\subsection{The optimal split at independence}\label{S-si:C12_transition}
At independence ($k=1/2$), an equal-size split that keeps the shares gives the classes $k_\pm=1/2\pm\varepsilon$ and the same squared consolidation $s=4\varepsilon^2$, hence the same indices. Averaging and taking absolute values therefore commute, and $J_{\rm strict}=J=|\chiA(s)|+|\chiB(s)|$ for every split, whatever the sign of~$\alpha\beta$. The optimum is as follows.
\begin{itemize}
\item In the agreement phase ($\alpha\beta>0$), $J(s)$ is increasing, and its minimum $|\alpha|+|\beta|$ is at $s=0$: the best option is not to split.
\item In the conflict phase ($\alpha\beta<0$, $|\alpha|\leq|\beta|$), the minimum is at $s=s^*$, where $\langle\chiA\rangle=0$, $\langle\chiB\rangle=(\beta^2-\alpha^2)/[\beta(1-\alpha^2)]$, and
\begin{equation}\label{S-eq:Jmin_C12}
J_{\min}=\frac{\beta^2-\alpha^2}{|\beta|\,(1-\alpha^2)}<|\beta|\leq|\alpha|+|\beta|.
\end{equation}
\end{itemize}

In the agreement phase with $\alpha,\beta>0$, both indices are positive and increasing in~$s$ (section~\ref{S-si:C12_index}), so $J=\chiA+\chiB$ is strictly increasing; with $\alpha,\beta<0$ the same holds for $J=-\chiA-\chiB$. In the conflict phase, take $\alpha>0>\beta$ without loss of generality. On $[0,s^*]$, $\chiA\geq0$ and $\chiB<0$, so
\[
J(s)=\chiA-\chiB=\frac{(\alpha-\beta)(1-s)}{1+\alpha\beta s},\qquad \frac{dJ}{ds}=-\frac{(\alpha-\beta)(1+\alpha\beta)}{(1+\alpha\beta s)^2}<0,
\]
since $\alpha-\beta>0$ and $1+\alpha\beta>0$. On $[s^*,1]$, $\chiA\leq0$ and $\chiB<0$, so
\[
J(s)=-\chiA-\chiB=\frac{-(\alpha+\beta)(1+s)}{1+\alpha\beta s},\qquad \frac{dJ}{ds}=\frac{-(\alpha+\beta)(1-\alpha\beta)}{(1+\alpha\beta s)^2}>0,
\]
since $\alpha+\beta<0$ and $1-\alpha\beta>1$. The unique minimum is therefore at~$s^*$, where $\chiA=0$ and $J=|\chiB(s^*)|$. Substituting $s^*=-\alpha/\beta$ gives $\chiB(s^*)=(\beta^2-\alpha^2)/[\beta(1-\alpha^2)]$, which has the sign of~$\beta$, and hence Eq.~\ref{S-eq:Jmin_C12}. The strict inequality holds because $(1-\alpha^2/\beta^2)/(1-\alpha^2)<1$ whenever $\alpha\neq0$ and $\beta^2<1$. For $\alpha=-0.2$ and $\beta=0.6$, $s^*=1/3$ and $J_{\min}=0.32/0.576\approx0.56$, against $0.80$ without a split.

\paragraph{Away from independence, in the agreement phase.} The agreement-phase result extends to schools away from independence ($k\neq1/2$), as stated in the main text. Not splitting remains exactly optimal for every~$k$ when $\alpha\beta\leq1/3$. When $\alpha\beta>1/3$, a split can lower~$J$, but only in strongly consolidated schools and by at most $0.12\%$ of~$J$.

By the symmetry $k\leftrightarrow1-k$ we take $k\geq1/2$ and write $u=2k-1$. A split with offset $\varepsilon\in[0,(1-u)/2]$ gives the two classes the signed consolidations $v=u\pm2\varepsilon$, whose squares are the squared consolidations $s_\pm$ of the classes. In the agreement phase, $\chiA$ and~$\chiB$ have the signs of $\alpha$ and~$\beta$ in every class (section~\ref{S-si:C12_index}), so the absolute values in~$J$ can be taken class by class, $J=J_{\rm strict}$, and
\begin{equation}\label{S-eq:J_agreement}
J(\varepsilon)=\tfrac12\bigl[G(u+2\varepsilon)+G(u-2\varepsilon)\bigr],\qquad
G(v)=T(v^2)=\frac{|\alpha+\beta|\,(1+v^2)}{1+\alpha\beta v^2},
\end{equation}
where $T(s)=|\chiA(s)|+|\chiB(s)|$ is the total segregation of a class (section~\ref{S-si:C12_strict}); here the two indices have the same sign, and $T$ is the absolute value of their sum. The undivided school has $J(0)=G(u)$, and $J(\varepsilon)$ is the average of~$G$ at two points whose midpoint is~$u$. With $c=\alpha\beta\in(0,1)$ and $w=v^2$,
\[
G''(v)=\frac{2\,|\alpha+\beta|\,(1-c)\,(1-3cw)}{(1+cw)^3}.
\]
If $c\leq1/3$, then $G''\geq0$ on $[-1,1]$, so $G$ is convex. The average of a convex function at two points is at least its value at their midpoint (Jensen's inequality), hence $J(\varepsilon)\geq J(0)$ for every feasible~$\varepsilon$: no split lowers~$J$, whatever the consolidation of the school.

If $c>1/3$, $G$ is concave where $v^2>1/(3c)$, and a split that moves one class into that region can lower~$J$. To bound the reduction, we computed $J(\varepsilon)$ from Eq.~\ref{S-eq:J_agreement}, which is exact in the model. The relative reduction $[J(0)-\min_\varepsilon J(\varepsilon)]/J(0)$ depends only on $c$ and~$k$, because the factor $|\alpha+\beta|$ cancels, so a grid of $c\in(1/3,1)$ and $k\in[1/2,1)$, both in steps of $0.0005$, with $2{,}001$ offsets for each~$k$, covers all pairs $(\alpha,\beta)$. The largest reduction is $0.12\%$ of~$J$, at $\alpha\beta\approx0.77$ and $k\approx0.89$, for the split that makes one class perfectly consolidated. On this grid, no reduction occurs when $(2k-1)^2\leq1/(3\alpha\beta)$: a split can help only in strongly consolidated schools.

\subsection{The optimal split judged class by class}\label{S-si:C12_strict}
Away from independence, a split with offset~$\varepsilon$ gives the classes $k_\pm=k\pm\varepsilon$ and squared consolidations $s_\pm=(2k_\pm-1)^2$, which now differ. Define the total segregation of a class,
\begin{equation}\label{S-eq:F_def}
T(s)=|\chiA(s)|+|\chiB(s)|,
\end{equation}
so that $J_{\rm strict}=\tfrac12[T(s_+)+T(s_-)]$: minimizing it over~$\varepsilon$ means choosing two points of the graph of~$T$ whose average is lowest.

\begin{figure}[tbp]
\centering
\includegraphics[width=0.65\textwidth]{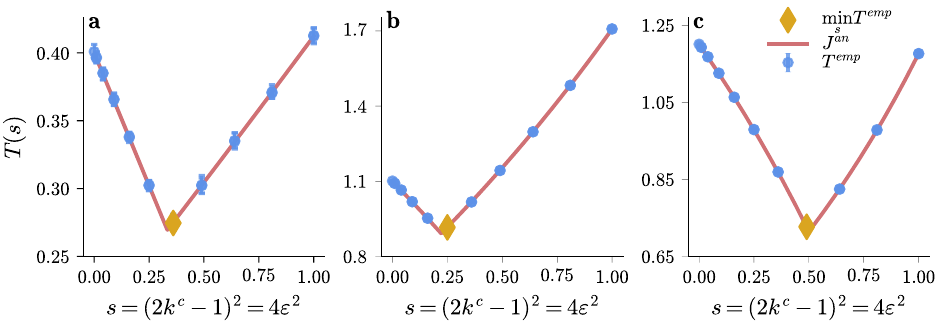}
\caption{\textbf{The total segregation of a class has a V-shape in the conflict phase.} $T(s)=|\chiA(s)|+|\chiB(s)|$ against the squared consolidation~$s$, for (a)~$\alpha=0.1$, $\beta=-0.3$; (b)~$\alpha=0.2$, $\beta=-0.9$; (c)~$\alpha=0.4$, $\beta=-0.8$. The left branch ($s<s^*$) is concave and the right branch ($s>s^*$) convex; the minimum is at $s^*=|\alpha|/|\beta|$. Lines: calculated, labeled $J^{an}$ in the legend; symbols: measured on networks simulated from the MAM, labeled $T^{emp}$; diamond: the smallest measured value, $\min_s T^{emp}$.}
\label{S-fig:F_shape}
\end{figure}

\paragraph{The shape of $T$.} In the conflict phase, with $\alpha>0>\beta$ and $|\beta|\geq|\alpha|$, $\chiB<0$ on $[0,1]$ and $\chiA$ changes sign at~$s^*$ (section~\ref{S-si:C12_index}), so
\begin{equation}\label{S-eq:F_branches}
T(s)=
\begin{cases}
\dfrac{(\alpha-\beta)(1-s)}{1+\alpha\beta s}, & s\leq s^*,\\[8pt]
\dfrac{-(\alpha+\beta)(1+s)}{1+\alpha\beta s}, & s\geq s^*,
\end{cases}
\end{equation}
and both branches equal $T(s^*)=(\beta^2-\alpha^2)/[|\beta|(1-\alpha^2)]$ at~$s^*$. The first derivatives were computed in section~\ref{S-si:C12_transition}: $T$ is strictly decreasing on $[0,s^*]$ and strictly increasing on $[s^*,1]$, a V-shape with its minimum at~$s^*$. The second derivatives,
\[
T''=\frac{2(\alpha-\beta)(1+\alpha\beta)\,\alpha\beta}{(1+\alpha\beta s)^3}<0\quad(s<s^*),\qquad
T''=\frac{2(\alpha+\beta)(1-\alpha\beta)\,\alpha\beta}{(1+\alpha\beta s)^3}>0\quad(s>s^*),
\]
show that the left branch is strictly concave and the right branch strictly convex (Fig.~\ref{S-fig:F_shape}). The V is generally asymmetric: the ratio of the one-sided slopes at~$s^*$,
\begin{equation}\label{S-eq:V_asymmetry}
\frac{|T'(s^{*-})|}{|T'(s^{*+})|}=\frac{(\alpha-\beta)(1+\alpha\beta)}{-(\alpha+\beta)(1-\alpha\beta)},
\end{equation}
equals~1 only when $\alpha=0$.

\paragraph{The straddling window.} Without a split, both classes sit at $s_0=(2k-1)^2$. The effect of a split depends on whether $s_0$ lies on the concave left branch ($s_0<s^*$) or on the convex right branch ($s_0\geq s^*$). The condition $s_0<s^*$ reads $|2k-1|<\sqrt{s^*}$, that is,
\begin{equation}\label{S-eq:cancellation_window}
k\in(k_{\rm low},k_{\rm high}),\qquad k_{\rm low}=\frac{1-\sqrt{s^*}}{2},\qquad k_{\rm high}=\frac{1+\sqrt{s^*}}{2},
\end{equation}
the straddling window of the main text. In the conflict phase, judged by~$J_{\rm strict}$, a split helps if and only if the school lies inside this window. Outside it, the best option is not to split. Inside it, with $u=|2k-1|$ and the largest feasible offset $\varepsilon_{\max}=\min(k,1-k)=(1-u)/2$, let
\[
\varepsilon_1=\frac{\sqrt{s^*}-u}{2},\qquad \varepsilon_2=\frac{\sqrt{s^*}+u}{2}.
\]
Both offsets place one class at~$s^*$; the other class sits at $\sigma=(2u-\sqrt{s^*})^2<s^*$ for~$\varepsilon_1$ and at $\tau=(2u+\sqrt{s^*})^2>s^*$ for~$\varepsilon_2$. $J_{\rm strict}$ is strictly decreasing for $\varepsilon\in[0,\varepsilon_1]$ and strictly increasing for $\varepsilon\geq\varepsilon_2$, so every optimal offset lies in $[\varepsilon_1,\bar\varepsilon]$, with $\bar\varepsilon=\min(\varepsilon_2,\varepsilon_{\max})$. Near $k=1/2$ the better end is~$\varepsilon_2$; farther from $k=1/2$ it is~$\varepsilon_1$, or $\varepsilon_{\max}$ when $\varepsilon_2$ is not feasible. At $k=1/2$ the two offsets coincide and both classes sit at~$s^*$, as in section~\ref{S-si:C12_transition}.

To derive these statements, use the symmetry $k\leftrightarrow1-k$ to take $k\geq1/2$, so $u=2k-1$. Let $r=\sqrt{s^*}$ and, for $v\in[-1,1]$, let $G(v)=T(v^2)$, an even function. Writing $x=u+2\varepsilon$, the two classes sit at $s_+=x^2$ and $s_-=(2u-x)^2$, so that $J_{\rm strict}=\tfrac12H(x)$ with $H(x)=G(x)+G(2u-x)$ and $x\in[u,1]$, where $x=1$ corresponds to~$\varepsilon_{\max}$. From Eq.~\ref{S-eq:F_branches}, with $c=\alpha\beta<0$ and $w=v^2$,
\begin{gather*}
G''(v)=\frac{2(\alpha-\beta)(1+c)(3cw-1)}{(1+cw)^3}<0\quad(|v|<r),\\
G''(v)=\frac{-2(\alpha+\beta)(1-c)(1-3cw)}{(1+cw)^3}>0\quad(|v|>r).
\end{gather*}
Hence $G$ is strictly concave on $[-r,r]$, where it is largest at $v=0$, and strictly convex on $[-1,-r]$ and on $[r,1]$, where it is decreasing and increasing, respectively; its minimum $T(s^*)$ is at $v=\pm r$.

\emph{Outside the window} ($u\geq r$). Let $x\in(u,1]$ and $y=2u-x$, so that $x>u\geq r$. If $y\geq r$, convexity of~$G$ on $[r,1]$ gives $G(x)+G(y)\geq2G(u)$. If $|y|<r$, then $G(y)>G(r)$; since $x+r>x+y=2u$, convexity and monotonicity of~$G$ on $[r,1]$ give $G(x)+G(r)\geq2G\bigl(\tfrac{x+r}{2}\bigr)\geq2G(u)$. If $y\leq-r$, then $G(y)=G(|y|)$ with $|y|\geq r$ and $x+|y|=2x-2u>2u$, so again $G(x)+G(|y|)\geq2G(u)$. In all cases $H(x)\geq H(u)$: the best offset is zero, and $J_{\rm strict}=J=T(u^2)$.

\emph{The aggregate objective outside the window.} No split lowers~$J$ either, so outside the window the two objectives have the same minimum, $T(u^2)$, reached without a split. With $\alpha>0>\beta$ and $g(v)=\chiA(v^2)$ as in section~\ref{S-si:C12_index}, $\langle\chiA\rangle=\tfrac12[g(u+2\varepsilon)+g(u-2\varepsilon)]$. (i)~By the concavity of~$g$, $\langle\chiA\rangle\leq g(u)=\chiA(u^2)\leq0$, the last inequality because $u^2\geq s^*$. With $D=\chiA(u^2)-\langle\chiA\rangle\geq0$, $|\langle\chiA\rangle|=|\chiA(u^2)|+D$: the split raises segregation on~$A$ by~$D$. (ii)~The pairs $(\chiA(s),\chiB(s))$ lie on a straight line of slope $m=\alpha(1-\beta^2)/[\beta(1-\alpha^2)]<0$ (section~\ref{S-si:C12_index}), so averaging over the two classes gives $\langle\chiB\rangle=\chiB(u^2)+m\,[\langle\chiA\rangle-\chiA(u^2)]=\chiB(u^2)+|m|D$. Since $\chiB<0$ in every class, $|\langle\chiB\rangle|=|\chiB(u^2)|-|m|D$: the split lowers segregation on~$B$ by~$|m|D$. (iii)~Adding the two, $J(\varepsilon)=J(0)+D\,(1-|m|)$. (iv)~Finally,
\[
|m|=\frac{\alpha}{1-\alpha^2}\cdot\frac{1-\beta^2}{|\beta|}\leq1,
\]
because the function $x\mapsto x/(1-x^2)$ is increasing on $[0,1)$ and $\alpha\leq|\beta|$, so that $\alpha/(1-\alpha^2)\leq|\beta|/(1-\beta^2)$. Hence $J(\varepsilon)\geq J(0)$ for every feasible offset, and $\min J=J(0)=\min J_{\rm strict}$.

\begin{figure}[tbp]
\centering
\includegraphics[width=0.7\textwidth]{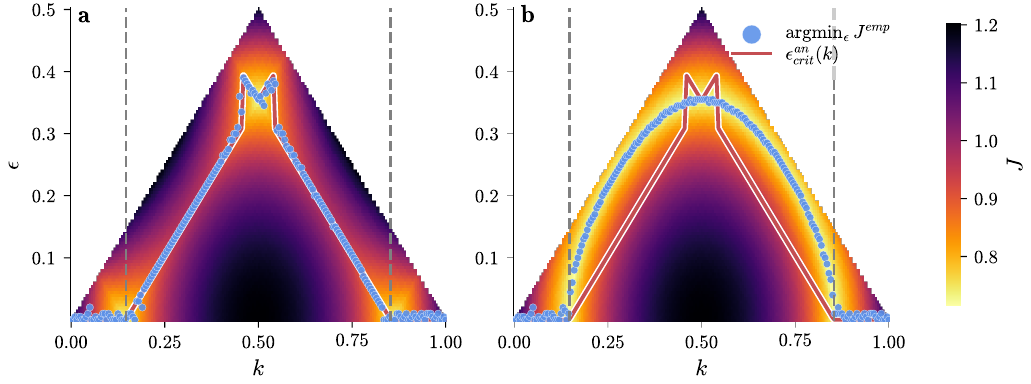}
\caption{\textbf{The class-by-class optimum and the aggregate optimum in the $(k,\varepsilon)$ plane.} Balanced school with $\alpha=0.4$, $\beta=-0.8$. Color: value of the objective; white: infeasible offsets ($\varepsilon>\min(k,1-k)$); gray dashes: the straddling window $(k_{\rm low},k_{\rm high})\approx(0.146,0.854)$. Red curve, $\varepsilon^{an}_{\rm crit}$ in the legend: the rule~\eqref{S-eq:Jstrict_comparison}, which uses $\varepsilon_1$ except in the narrow band where $\varepsilon_2$ is feasible, $(1+\sqrt{s^*})/4\leq k\leq(3-\sqrt{s^*})/4$, and gives a lower~$J_{\rm strict}$. Blue circles, $\mathrm{argmin}_\varepsilon J^{emp}$ in the legend: numerical minimizers over~$\varepsilon$. (a)~$J_{\rm strict}$: the rule follows the numerical minimum throughout the window; the small kink near $k\approx0.5$ is the switch between $\varepsilon_2$ and~$\varepsilon_1$. (b)~$J$, on the same color scale: the offset that minimizes~$J$ lies far beyond the class-by-class optimum, where segregation on~$A$ cancels between the classes; the deeper valley of~$J$ inside the window is the signature of the cancellation trap.}
\label{S-fig:Jstrict_heatmap}
\end{figure}

\emph{Inside the window} ($u<r$). (a)~For $x\in[u,r]$, that is, $\varepsilon\in[0,\varepsilon_1]$, both $x$ and $2u-x$ lie in $[-r,r]$, so $H$ is strictly concave there; since $H'(u)=0$, $H$ is strictly decreasing on $[u,r]$, and any split with $0<\varepsilon\leq\varepsilon_1$ strictly lowers~$J_{\rm strict}$. (b)~For $x\geq2u+r$, that is, $\varepsilon\geq\varepsilon_2$, we have $x>r$ and $2u-x\leq-r$, so $H'(x)=G'(x)-G'(2u-x)>0$. (c)~Every minimizer therefore lies in $[\varepsilon_1,\bar\varepsilon]$, and $\varepsilon_2$ is feasible if and only if $2u+r\leq1$. The strict objective at the ends of this interval is
\begin{align}
J_{\rm strict}(\varepsilon_1)&=\frac{|\alpha-\beta|\,(1-\sigma)}{2(1+\alpha\beta\sigma)}+\frac{T(s^*)}{2},\label{S-eq:Jstrict_opt_e1}\\
J_{\rm strict}(\varepsilon_2)&=\frac{|\alpha+\beta|\,(1+\tau)}{2(1+\alpha\beta\tau)}+\frac{T(s^*)}{2},\label{S-eq:Jstrict_opt_e2}\\
J_{\rm strict}(\varepsilon_{\max})&=\frac{|\alpha+\beta|}{1+\alpha\beta}+\frac{|\alpha-\beta|\,(1-\sigma_m)}{2(1+\alpha\beta\sigma_m)},\label{S-eq:Jstrict_opt_emax}
\end{align}
where at $\varepsilon_{\max}$ (when $2u+r>1$) the classes sit at~1 and at $\sigma_m=(1-2u)^2<s^*$. (d)~For small~$u$, $\sigma\approx s^*-4ru$ and $\tau\approx s^*+4ru$, so $T(\sigma)-T(\tau)\approx4ru\,\bigl(|T'(s^{*-})|-T'(s^{*+})\bigr)>0$ by Eq.~\ref{S-eq:V_asymmetry}: near $k=1/2$ the end~$\varepsilon_2$ is the better one. (e)~On $[\varepsilon_1,\bar\varepsilon]$, $H$ is the sum of a convex and a concave term and need not be concave. When it is concave, the better end is optimal, which gives the rule
\begin{equation}\label{S-eq:Jstrict_comparison}
\varepsilon_{\rm opt}(k)=
\begin{cases}
\varepsilon_2 & \text{if } \varepsilon_2\leq\varepsilon_{\max} \text{ and } T(\tau)<T(\sigma),\\
\varepsilon_{\max} & \text{if } \varepsilon_2>\varepsilon_{\max} \text{ and } J_{\rm strict}(\varepsilon_{\max})<J_{\rm strict}(\varepsilon_1),\\
\varepsilon_1 & \text{otherwise}.
\end{cases}
\end{equation}
Otherwise an interior minimum is possible. In a numerical search over a dense grid of $(\alpha,\beta,k)$, the optimum lay strictly inside $(\varepsilon_1,\bar\varepsilon)$ only for strong preferences and in a narrow band of~$k$ around the value at which the better end switches from $\varepsilon_2$ to~$\varepsilon_1$; in all the cases examined, the rule exceeded the minimum of~$J_{\rm strict}$ by less than $0.01$, and for the parameters of Fig.~\ref{S-fig:Jstrict_heatmap} by less than~$10^{-4}$.

Fig.~\ref{S-fig:Jstrict_heatmap} compares the rule with a numerical minimization over~$\varepsilon$, for both objectives.

\subsection{The cancellation trap}\label{S-si:C12_trap}
The optima of~$J$ and~$J_{\rm strict}$ agree outside the straddling window (section~\ref{S-si:C12_strict}) but differ inside it. The difference comes entirely from attribute~$A$: with $\alpha>0>\beta$ and $|\alpha|\leq|\beta|$, $\chiB(s)<0$ for all $s\in[0,1]$, so $\langle|\chiB|\rangle=|\langle\chiB\rangle|$ for every split. Inside the window and for $k\neq1/2$, $J$ can reach its floor by perfect cancellation: an offset at which $\langle\chiA\rangle=0$ while $|\chiA(s_\pm)|>0$. This zero is necessarily a cancellation, since $\chiA=0$ in both classes would require $s_+=s_-=s^*$, which is possible only at $k=1/2$. Because the attainable pairs lie on a straight segment (section~\ref{S-si:C12_index}), $\langle\chiA\rangle=0$ fixes $\langle\chiB\rangle=\chiB(s^*)$, so the floor of~$J$ is $|\chiB(s^*)|$ and does not depend on~$k$. Outside the window, $\langle\chiA\rangle\leq\chiA(s_0)\leq0$ for every split (section~\ref{S-si:C12_strict}), so a cancellation is possible only inside it.

\paragraph{When the floor can be reached.} Whether the floor can be reached is a question of feasibility: $\langle\chiA\rangle$ must change sign as $\varepsilon$ runs from 0 to $\varepsilon_{\max}=\min(k,1-k)$. At $\varepsilon=0$ it is the index of the undivided school, $\chiA(s_0)$ with $s_0=(2k-1)^2$, which is positive because $s_0<s^*$ inside the window; it varies continuously with~$\varepsilon$ and decreases as $\varepsilon$ grows (section~\ref{S-si:C12_index}). It therefore has a zero on $[0,\varepsilon_{\max}]$, and the floor is reached, if and only if $\langle\chiA\rangle\leq0$ at~$\varepsilon_{\max}$. At this offset one class is perfectly consolidated ($s=1$); for $k\geq1/2$ the other has $k-\varepsilon_{\max}=2k-1$, hence $s=(4k-3)^2$, and the condition reads $\chiA(1)+\chiA\big((4k-3)^2\big)\leq0$. Since $\chiA$ decreases with~$s$, the second term is at most $\chiA(0)=\alpha$, with equality at $k=3/4$, where the other class is at independence; by the symmetry $k\leftrightarrow1-k$, the same holds at $k=1/4$. The condition is therefore hardest to meet at $k=1/4$ and $k=3/4$, and the floor is reachable at every~$k$ in the window if and only if $\alpha+\chiA(1)\leq0$, that is, $s^*\leq(1+\alpha^2)/2$; when this fails, $s^*>1/2$, so $k=1/4$ and $k=3/4$ lie inside the window. The condition holds for the parameters of Fig.~\ref{fig:trap} of the main text ($\alpha=0.4$, $\beta=-0.8$), where $\alpha+\chiA(1)=0.4-0.59<0$, but not, for instance, for $\alpha=0.4$ and $\beta=-0.6$, where $\alpha+\chiA(1)=0.4-0.26>0$: the floor is then out of reach for $0.12<k<0.38$ and $0.62<k<0.88$, about 60\% of the window, and reachable near the edges of the window and near independence.

\paragraph{When every class is more segregated.} When perfect cancellation is feasible, the split that minimizes~$J$ gives the two classes indices of~$A$ of opposite sign and equal size, $|\chiA(s_+)|=|\chiA(s_-)|>0$. This common value is not always larger than the value $|\chiA(s_0)|$ of the undivided school: close to $k=1/2$ it is smaller (for $\alpha=0.4$, $\beta=-0.8$ and $k=0.506$, the class value is $0.016$ against $0.40$ for the school), while close to the edges of the window it is larger. With $u=2k-1$, both classes are more segregated on~$A$ than the undivided school if and only if $|u|>1/3$ or $\chiA(u^2)+\chiA(9u^2)$ has the sign of~$\beta$; for $\alpha=0.4$ and $\beta=-0.8$, if and only if $|2k-1|>0.30$, the threshold quoted in the main text.

To derive this condition, take $\alpha>0>\beta$ (for $\alpha<0<\beta$ both indices change sign) and, by the symmetry $k\leftrightarrow1-k$, $k>1/2$. Then $u=2k-1\in(0,\sqrt{s^*})$, $\delta=\chiA(s_0)>0$ and $\varepsilon_{\max}=(1-u)/2$; the classes have signed consolidations $v_\pm=u\pm2\varepsilon$ and $s_\pm=v_\pm^2$, and $\chiA$ is strictly decreasing in~$s$. Let $\varepsilon_J$ be the offset at which $\langle\chiA\rangle=0$. (i)~Both classes are more segregated on~$A$ than the undivided school if and only if $\varepsilon_J<u$. If $\varepsilon_J<u$, then $|v_-|<u$, so $s_-<s_0$ and $\chiA(s_-)>\delta$; hence $|\chiA(s_\pm)|>\delta$. If $\varepsilon_J\geq u$, then $|v_-|\geq u$, so $s_-\geq s_0$ and $\chiA(s_-)\leq\delta$; moreover $\chiA(s_-)>0$, because $s_-<s_+$ and the two classes have opposite signs, so $|\chiA(s_\pm)|=\chiA(s_-)\leq\delta$. (ii)~Since $\langle\chiA\rangle$ decreases strictly as $\varepsilon$ grows (section~\ref{S-si:C12_index}) and vanishes only at~$\varepsilon_J$: if $u>\varepsilon_{\max}$, that is, $u>1/3$, then $\varepsilon_J\leq\varepsilon_{\max}<u$. Otherwise $\varepsilon_J<u$ if and only if $\langle\chiA\rangle<0$ at $\varepsilon=u$, where the two classes sit at $s_-=u^2$ and $s_+=9u^2$, that is, if and only if $\chiA(u^2)+\chiA(9u^2)<0$. (iii)~Both $\chiA(u^2)$ and $\chiA(9u^2)$ decrease as $u$ grows, so the condition of (ii) holds exactly for $u$ above a threshold $u_{\rm crit}$, and $k_{\rm crit}=(1-u_{\rm crit})/2$. The threshold lies inside the window: as $u\to\sqrt{s^*}$, $\chiA(u^2)\to0$, while $\chiA(9u^2)<0$ if $9u^2\leq1$ (otherwise $u>1/3$). For $\alpha=0.4$ and $\beta=-0.8$, $u_{\rm crit}=0.30$.

\subsection{When both classes improve}\label{S-si:C12_both}
\begin{figure}[tbp]
\centering
\includegraphics[width=0.65\textwidth]{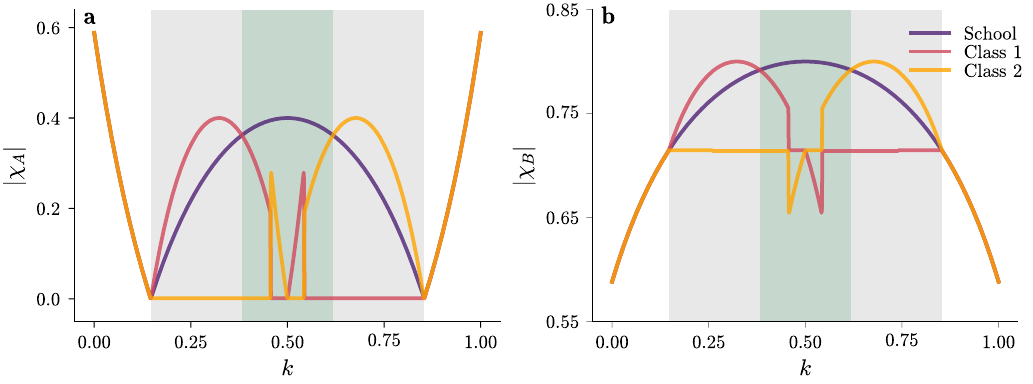}
\caption{\textbf{Which classes improve under the class-by-class optimum.} Class indices under the split that minimizes~$J_{\rm strict}$ in a balanced school ($\alpha=0.4$, $\beta=-0.8$, $s^*=1/2$), against the consolidation~$k$ of the school; all lines are calculated. Gray band: straddling window; green band: inner window, where both classes are less segregated than the school on both attributes. (a)~$|\chiA|$. Outside the straddling window the optimum is not to split, so the classes equal the school. Inside it, the optimal split places one class at $\chiA(s^*)=0$ and the other at $\sigma$ or~$\tau$, depending on the distance of $k$ from~$1/2$; near $k=1/2$, where the optimal offset switches between $\varepsilon_1$ and~$\varepsilon_2$, the class placed at~$s^*$ changes, hence the jumps. (b)~The same for $|\chiB|$; at these parameters $\chiB$ is negative for every~$k$.}
\label{S-fig:chiA_vs_k_benefit}
\end{figure}

This subsection finds the consolidations~$k$ for which the split that minimizes~$J_{\rm strict}$ lowers segregation in both classes and on both attributes, compared with the undivided school; the main text quotes the threshold for $\alpha=0.4$ and $\beta=-0.8$. Outside the straddling window this cannot happen, since the optimum is not to split. Inside it, the class placed at~$s^*$ always has lower segregation than the undivided school on both attributes. Whether the other class does too depends on~$k$. At the end~$\varepsilon_1$, the other class sits at $\sigma=(2u-\sqrt{s^*})^2$, on the left branch, where $\chiA$ decreases and $|\chiB|$ decreases with~$s$; it improves on both attributes if and only if $\sigma>s_0=u^2$, that is, $|2u-\sqrt{s^*}|>u$, which for $u<\sqrt{s^*}$ means $u<\sqrt{s^*}/3$. This defines the inner window
\[
\tfrac12-\tfrac{\sqrt{s^*}}{6}<k<\tfrac12+\tfrac{\sqrt{s^*}}{6}.
\]
For the parameters of Fig.~\ref{S-fig:chiA_vs_k_benefit}, both classes are less segregated than the undivided school on both attributes exactly when $k$ lies in the inner window; farther from $k=1/2$, the other class is more segregated than the undivided school on both attributes. The inner window is not a general condition, because the optimum may be at the end~$\varepsilon_2$, where the other class sits at $\tau>s^*$ and $\chiA(\tau)<0$: for $\alpha=0.1$, $\beta=-0.3$ and $k=0.57$, the optimum is $\varepsilon_2$ and $k$ lies in the inner window, yet the class at~$\tau$ has $|\chiA|=0.12$, against $0.094$ in the undivided school.

\section{Data and estimation}\label{S-si:data}
This section describes the friendship networks (section~\ref{S-si:cils4eu}), the estimation of the preferences and the comparison of the models (section~\ref{S-si:inference}), the networks simulated from the estimated models (section~\ref{S-si:replicas}), and the tests of the splitting results on the data (section~\ref{S-si:splittests}). It supports the subsection of Results ``A single-attribute model misattributes segregation in empirical school networks'', the data tests of the subsections on splitting, and the Materials and Methods paragraphs ``Data'', ``Model estimation'' and ``Class-splitting experiments''.

\subsection{Network data}\label{S-si:cils4eu}
The networks come from the Children of Immigrants Longitudinal Survey in Four European Countries (CILS4EU) \citep{kalter2016cils4eu}, a study of the intergenerational integration of children of immigrants in England, Germany, the Netherlands and Sweden. The first wave was conducted in 2010; we use the second wave, collected in 2011 and 2012. Students completed questionnaires on, among other topics, their school performance, family background and socioeconomic status, and named the classmates with whom they had several kinds of relation. We build the networks from these nominations.

The dataset identifies the class and the school of each student. Since nominations were collected among classmates, we take the class as the pool of possible friends of a student and build one network per class. We first checked whether classes of the same school ever mix, that is, whether students named students of another class of their school. In most schools, classes form separate components. In the few schools where classes do mix (for example, units 1045, 1077 and 1096 in England), we build the network at the school level, so that no friendship is cut by a class boundary. Each nomination points from the respondent to the classmate named, so the networks are directed; when the same pair appears in several items, the tie is kept once.

We use the following Wave~2 variables of the reduced-version files (CILS4EU codebook, wave~2, v2.3.0, Mannheim University, 2016). Identifiers: \texttt{youthid}, the class and school survey units \texttt{y2\_svyunit\_c} and \texttt{y2\_svyunit\_s}, and \texttt{country}. Attributes, from the youth main questionnaire: sex (\texttt{y2\_sex}) and the countries of birth of the respondent, the mother and the father (\texttt{y2\_cobRV}, \texttt{y2\_cobmRV}, \texttt{y2\_cobfRV}). Ties, from the youth classmates questionnaire: best friends in class (\texttt{y2\_bfs\_*}, up to five nominations), spending time together outside school (\texttt{y2\_tos\_*}), and doing homework together (\texttt{y2\_hw\_*}).

Each student has two binary attributes: sex (male or female) and country of origin (migration background). Following \citet{dollmann2016examining} and \citet{boller2025friendship}, a student is classified as of foreign origin if the student or either parent was born outside the survey country, giving precedence to the mother's country when the parents' countries differ; a student is classified as of survey-country origin only when the student and both parents were born in the survey country. Students with a missing attribute or survey unit are removed. In each network, attribute~$A$ is the one whose minority group is smaller, so that $A$ can be either country of origin or sex.

We keep the networks that satisfy four criteria. First, a network must contain at least 20 students, so that all combinations of the two attributes are represented. Second, both groups of both attributes must be present, since a network with a single group on one attribute carries little information on mixing between groups. Third, at least one student of each of the four types $(A_a,B_b)$ must be present, so that all intersectional groups are represented. Fourth, the network must form a single connected component, so that measures that depend on the whole network are well defined. The final sample has 139 networks; Table~\ref{S-tab:allnets} summarizes them by country.

\begin{table}[tbp]\footnotesize
\centering
\begin{tabular}{lcccccccc}
\toprule
Country & \makecell{Number of \\ networks} & $N$ range & \makecell{$N$\\median} & \makecell{Density\\range} & \makecell{Density\\median} & \makecell{Min degree\\range} & \makecell{Max degree\\range} & \makecell{Mean degree\\range} \\
\midrule
EN & 23 & [20, 55] & 35 & [0.08, 0.32] & 0.14 & [1, 6] & [10, 33] & [6, 16] \\
GE & 33 & [20, 29] & 22 & [0.17, 0.59] & 0.35 & [2, 12] & [14, 42] & [9, 23] \\
NL & 29 & [20, 26] & 22 & [0.12, 0.42] & 0.21 & [1, 8] & [10, 35] & [6, 21] \\
SW & 54 & [20, 27] & 22 & [0.17, 0.47] & 0.25 & [1, 10] & [13, 44] & [8, 21] \\
\bottomrule
\end{tabular}
\caption{\textbf{The 139 school-class networks, by country.} Each network is a class (in a few cases a school) and each node a student. Ranges are taken across the networks of a country. Columns, from left to right: country code (England, Germany, the Netherlands, Sweden); number of networks; range and median of the number of nodes~$N$; range and median of the density, the number of ties out of the about $N^2$ possible; ranges of the minimum, maximum and mean degree.}
\label{S-tab:allnets}
\end{table}

The most widely used school friendship dataset in the segregation literature is Add Health \citep{harris2013add}, which covers schools in the United States from 1994. We use CILS4EU instead for two reasons. First, European school systems are organized in classes to which an administrator assigns the students, so the class splits studied in the main text are realistic interventions. Second, CILS4EU is more recent and records a rich set of attributes for the students of each class.

\subsection{Estimation and model comparison}\label{S-si:inference}
This subsection gives the models, the density parameter, the likelihood and the tests that Materials and Methods summarizes. We fit three models to each network. They are nested, SAM~$\subset$~MAM~$\subset$~uMAM: the SAM is the MAM with a neutral preference over one of the two attributes, and the MAM is the uMAM with symmetric, row-normalized preference matrices; the uMAM is the unconstrained model of \citet{martin2026simple}, called MAPS-AND by its authors (section~\ref{S-si:preferences}). We fit the SAM twice, with country of origin and with sex as its attribute; the main text reports the first.

The tie probabilities of the MAM and of the SAM are multiplied by a global density parameter~$\rho$, which does not depend on the attributes; in the uMAM, the free entries of the preference matrices set the density. The density parameter sets the overall density of ties and has no effect on the Coleman index, which compares fractions of ties, but it is needed for a fair comparison of the models, for two reasons. First, the uMAM can vary every entry of its preference matrices independently, whereas the MAM constrains them; without a density parameter, the MAM could not reproduce the low density of very sparse networks, and it would fit worse for this reason alone. Second, a network formed through a preference over~$A$ alone is generated by the MAM with the neutral matrix $H^B=\tfrac12\bigl(\begin{smallmatrix}1&1\\1&1\end{smallmatrix}\bigr)$, since the preference matrices are symmetric and row-normalized, and by the SAM with $H^A$ alone, so the SAM would generate a network twice as dense. With the density parameter, the tie probability is proportional to $\rho\,H^A_{A(i),A(j)}H^B_{B(i),B(j)}$ in the MAM and to $\rho\,H^X_{X(i),X(j)}$ in the SAM fitted on attribute~$X$. The density parameter is a scale factor, not a density: values of~$\rho$ above one are possible, as long as every tie probability stays below one (Table~\ref{S-tab:replicas}).

All parameters are estimated with a procedure close to that of \citet{martin2026simple}: we maximize the binomial log-likelihood of the observed numbers of ties between the four types $(A_a,B_b)$ of the two attributes (for instance, foreign origin and female), given the number of students of each type,
\begin{equation}\label{S-eq:loglik}
\ell=\sum_{t,t'}\Bigl[m_{tt'}\ln p_{tt'}+\bigl(M_{tt'}-m_{tt'}\bigr)\ln\bigl(1-p_{tt'}\bigr)\Bigr],
\end{equation}
where $t$ and~$t'$ run over the four types, $m_{tt'}$ is the number of ties from students of type~$t$ to students of type~$t'$, $M_{tt'}$ is the number of possible directed ties from students of type~$t$ to students of type~$t'$, given the composition of the class, and $p_{tt'}$ is the probability, in the model, of a tie from a student of type~$t$ to a student of type~$t'$, including the density parameter in the MAM and in the SAM.

Since all three are fitted by maximum likelihood on the same table of tie counts, their likelihoods are directly comparable, and we compare each nested pair with a likelihood-ratio test. The statistic $\mathrm{LR}=2\,(\log\hat L_{\rm larger}-\log\hat L_{\rm smaller})$ is, under the null hypothesis that the smaller model is correct, asymptotically distributed as $\chi^2_{\Delta k}$, where $\Delta k$ is the difference in the number of free parameters, two for the SAM (one preference and~$\rho$), three for the MAM ($\alpha$, $\beta$ and~$\rho$), and seven for the uMAM, whose eight entries are defined up to a factor that one matrix can pass to the other: $\Delta k=1$ for the MAM against the SAM and $\Delta k=4$ for the uMAM against the MAM. Rejecting the null hypothesis of the MAM-against-SAM test ($p<0.05$) means that the second attribute significantly improves the fit. Failing to reject the null hypothesis of the uMAM-against-MAM test ($p\geq0.05$) means that the extra flexibility of the unconstrained model does not significantly improve the fit, so that the MAM describes the data adequately.

All three models converge in all 139 networks. The MAM outperforms the SAM fitted on country of origin in all 139 networks, and the SAM fitted on sex in 74 of them. The constraints of the MAM are what make the theory tractable, and relaxing them gives the uMAM, which has more free parameters and can be expected to fit at least as well. The same trade-off between tractability and flexibility appears in the work of \citet{martin2025intersectional, martin2026simple}, who use a constrained version of their multiplicative model for analytical results and an unconstrained one for empirical validation. We therefore identify the networks in which the MAM performs comparably to the uMAM: the MAM is not rejected against the uMAM in 69 networks (49.6\%). The empirical analyses that use the estimated preferences, namely the simulated networks of section~\ref{S-si:replicas} and the tests of the splitting results, are restricted to these 69 networks, so that the estimates they rely on are as reliable as possible. The result that the MAM outperforms the SAM fitted on country of origin holds on the whole sample and is not affected by the restriction.

\subsection{Simulated networks}\label{S-si:replicas}
\begin{figure}[tbp]
\centering
\includegraphics[width=0.45\textwidth]{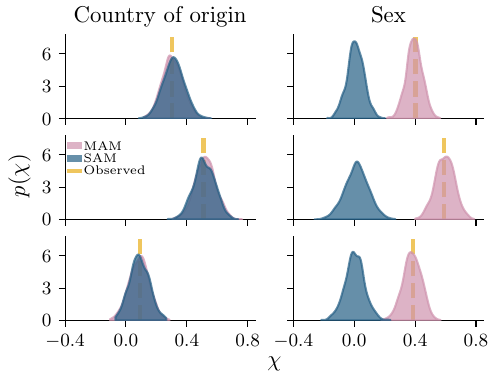}
\caption{\textbf{The single-attribute model reproduces segregation by origin but not by sex.} Coleman index of country of origin (left) and of sex (right) measured on 1,000 networks simulated from the MAM (pink) and from the SAM fitted on country of origin (blue), and on the observed network (dashed line), for three classes (rows; Table~\ref{S-tab:replicas}). Both models reproduce segregation by origin, which the SAM is fitted to; the SAM produces almost no segregation by sex, whereas the MAM reproduces it.}
\label{S-fig:replicas_origin}
\end{figure}

\begin{figure}[tbp]
\centering
\includegraphics[height=2.3in]{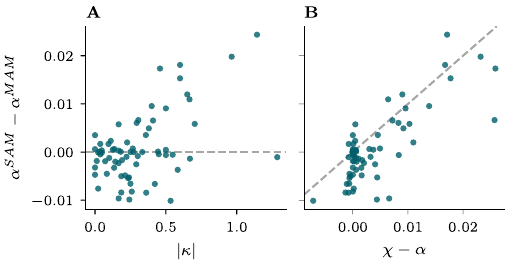}\hfill
\includegraphics[height=2.3in]{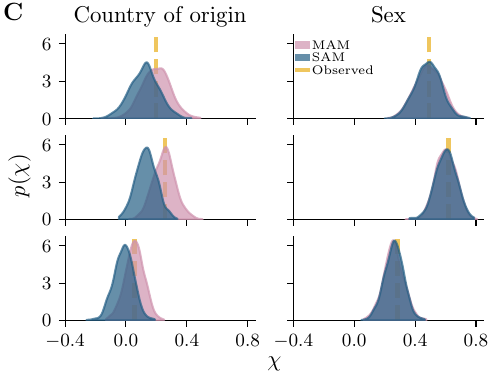}
\caption{\textbf{The single-attribute model fitted on sex.} (A)~Difference between the preferences over sex estimated with the SAM and with the MAM, $\alpha^{\rm SAM}-\alpha^{\rm MAM}$, against the absolute normalized consolidation $|\kappa|=|k-\fB|/\fB$, for the 69 networks in which the MAM fits as well as the uMAM; dashed line: zero. (B)~The same difference against the induced segregation $\chi-\alpha$ calculated from the closed form; dashed line: the diagonal. (C)~Coleman index of country of origin (left) and of sex (right) measured on 1,000 networks simulated from the MAM (pink) and from the SAM fitted on sex (blue), and on the observed network (dashed line), for three networks (rows; Table~\ref{S-tab:replicas}).}
\label{S-fig:replicas_sex}
\end{figure}

For the networks shown in Fig.~\ref{S-fig:replicas_origin} and in Fig.~\ref{S-fig:replicas_sex}C, we simulate synthetic networks from the observed composition~$F$ and the estimated preferences and density: students are assigned to the four types in the observed numbers, and directed ties are drawn independently with the estimated tie probabilities, density included (as in Materials and Methods, ``Simulations''). We generate 1,000 independent networks per class from the MAM and from the SAM, and measure the Coleman index of each attribute on every simulated network. The resulting distributions are compared, attribute by attribute, with the index measured on the observed network.

The networks are chosen by their $B$-imbalances. By Eq.~(\ref{eq:chiA_minus_alpha}) of the main text, induced segregation grows with the difference $\Delta=\delta_0-\delta_1$ between the $B$-imbalances of the two $A$-groups. Networks in which $\delta_0$ and~$\delta_1$ are large and of opposite signs, so that $\Delta$ is large, should therefore show the strongest effects of the second attribute. Table~\ref{S-tab:replicas} lists them.

\paragraph{The SAM fitted on sex.} Fig.~\ref{S-fig:replicas_sex} repeats the comparison of Fig.~\ref{fig:data}A of the main text and Fig.~\ref{S-fig:replicas_origin} with the SAM fitted on sex instead of country of origin. The MAM outperforms this SAM in 74 of the 139 networks, and in 34 of the 69 used for the simulations. The SAM fitted on sex inflates the preference over sex only slightly (panel A; differences of at most about 0.025, against about 0.2 for country of origin in Fig.~\ref{fig:data}A of the main text), and the inflation again grows with consolidation. Panel B compares it with the induced segregation calculated from the closed form. In the simulated networks (panel C), both models reproduce the segregation by sex measured on the observed networks, while the SAM produces less segregation by country of origin than observed: the mirror image of Fig.~\ref{S-fig:replicas_origin}.

\begin{table}[tbp]\footnotesize
\centering
\setlength{\tabcolsep}{4pt}
\begin{tabular}{cccccccccccccc}
\toprule
Unit & Country & \makecell{SAM\\attribute} & $N$ & $\fA$ & $\fB$ & $k$ & $\delta_0$ & $\delta_1$ & $h_A^{\rm MAM}$ & $h_B^{\rm MAM}$ & $h^{\rm SAM}$ & $\rho^{\rm MAM}$ & $\rho^{\rm SAM}$ \\
\midrule
211702 & GE & Origin & 22 & 0.23 & 0.50 & 0.20 & $-0.14$ & 0.14 & 0.65 & 0.70 & 0.66 & 1.45 & 0.71 \\
411401 & SW & Origin & 25 & 0.48 & 0.48 & 0.33 & $-0.16$ & 0.12 & 0.76 & 0.80 & 0.76 & 0.87 & 0.43 \\
104101 & EN & Origin & 42 & 0.41 & 0.48 & 0.65 & 0.12 & $-0.17$ & 0.53 & 0.69 & 0.55 & 0.49 & 0.25 \\
400301 & SW & Sex & 22 & 0.46 & 0.50 & 0.20 & $-0.27$ & 0.27 & 0.74 & 0.54 & 0.75 & 0.73 & 0.37 \\
411402 & SW & Sex & 22 & 0.32 & 0.36 & 0.71 & 0.14 & $-0.41$ & 0.80 & 0.56 & 0.81 & 1.06 & 0.54 \\
203302 & GE & Sex & 20 & 0.35 & 0.50 & 0.57 & 0.05 & $-0.05$ & 0.64 & 0.53 & 0.64 & 1.51 & 0.76 \\
\bottomrule
\end{tabular}
\caption{\textbf{Classes used for the simulated networks.} Survey unit, country, attribute of the SAM (country of origin or sex), number of students~$N$, composition $(\fA,\fB,k)$, $B$-imbalances $\delta_0$ and~$\delta_1$, estimated preferences, written as the diagonal entries $h=(1+\alpha)/2$ of the preference matrices ($h_A^{\rm MAM}$ and $h_B^{\rm MAM}$ for the MAM, $h^{\rm SAM}$ for the SAM), and estimated density parameters $\rho^{\rm MAM}$ and~$\rho^{\rm SAM}$. Rows with the SAM fitted on country of origin refer to Fig.~\ref{S-fig:replicas_origin}, rows with the SAM fitted on sex to Fig.~\ref{S-fig:replicas_sex}C, in order of appearance.}
\label{S-tab:replicas}
\end{table}

\subsection{Tests of the splitting results}\label{S-si:splittests}
This subsection describes how the splitting results of the main text are tested on the empirical networks. Attribute~$A$ is the one whose minority group is smaller, so it can be either country of origin or sex. Each class is described by its composition $(\fA,\fB,k)$ and by the preferences $(\alpha,\beta)$ estimated with the MAM, and only the 69 networks in which the MAM fits about as well as the uMAM are used (section~\ref{S-si:inference}).

\paragraph{From a target composition to a split.} Each empirical network is a single class, so we split classes into two subclasses rather than schools into classes. A splitting rule prescribes a target composition for subclass~1, that is, the number $C_1[a,b]$ of students of each type $(A_a,B_b)$; the remaining students form subclass~2. Within each type, the students assigned to subclass~1 are drawn at random, and every split is repeated 20 times. We keep only the ties within each subclass, compute $\chiA$ and~$\chiB$ in each subclass from the observed ties, and record the size-weighted averages $\langle\chi_A\rangle$, $\langle\chi_B\rangle$ and the aggregate objective~$J$. This assumes, as the model does, that the ties within a class do not depend on who else is in it; it does not capture how friendships would re-form after an actual reassignment. Classes with an odd number of students are split into two subclasses whose sizes differ by one. Subclasses have between 10 and 26 students.

\begin{figure}[tbp]
\centering
\includegraphics[width=0.5\textwidth]{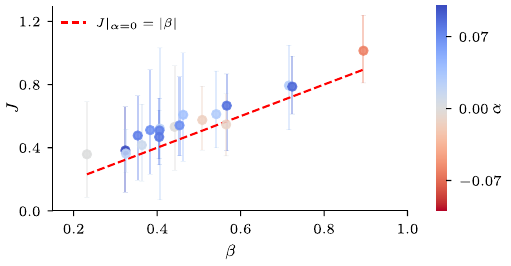}
\caption{\textbf{The ESDS splits of the classes with $\alpha\approx0$ reach the floor of the aggregate objective.} Aggregate objective~$J$ of the ESDS splits against the estimated~$\beta$, for the 19 CILS4EU classes with $|\alpha|\leq0.1$ in the region where the ESDS exists (Table~\ref{S-tab:esdsnets}); color: estimated~$\alpha$; dashed line: the floor~$|\beta|$. $J$ lies close to the floor, within the error bars; it combines the two averages shown in Fig.~\ref{fig:alpha0}C--D of the main text.}
\label{S-fig:esds_J}
\end{figure}

\paragraph{One neutral preference.} Since $\alpha=0$ is never attained exactly, we select the 20 classes with $|\alpha|\leq0.1$, about 29\% of the 69. To check the existence condition~\eqref{S-eq:rank1_condition} of the ESDS, for each value of~$\fA$ in the data we sweep a dense grid of $(\fB,k)$ and, at each grid point, the one-parameter family of candidate splits~\eqref{S-eq:rank1_param}; a grid point passes if some candidate gives two valid population matrices whose average is~$F$. Fig.~\ref{fig:esdsmap} of the main text shows the resulting map with the selected classes: all but one lie in the region where the ESDS exists. These 19 classes are used to test the ESDS, and Table~\ref{S-tab:esdsnets} lists them. For the ESDS, the targets $C_1[a,b]$ are taken from the family~\eqref{S-eq:rank1_param}, and the results are pooled over all admissible targets and repetitions (Fig.~\ref{fig:alpha0}C--D of the main text, and Fig.~\ref{S-fig:esds_J} for the aggregate objective).

\begin{table}[tbp]
\centering
\footnotesize
\begin{tabular}{lccccccc}
\toprule
Unit & Country & $N$ & $\fA$ & $\fB$ & $k$ & $\alpha$ & $\beta$ \\
\midrule
104101 & EN & 42 & 0.41 & 0.48 & 0.65 & 0.07 & 0.38 \\
213201 & GE & 23 & 0.26 & 0.39 & 0.67 & 0.10 & 0.32 \\
300437 & NL & 24 & 0.17 & 0.50 & 0.25 & 0.02 & 0.41 \\
408102 & SW & 26 & 0.23 & 0.46 & 0.67 & 0.03 & 0.72 \\
102501 & EN & 38 & 0.21 & 0.50 & 0.62 & 0.00 & 0.44 \\
412302 & SW & 22 & 0.14 & 0.50 & 0.33 & 0.00 & 0.23 \\
412602 & SW & 21 & 0.19 & 0.33 & 0.25 & 0.07 & 0.35 \\
400302 & SW & 20 & 0.35 & 0.45 & 0.43 & 0.03 & 0.54 \\
211001 & GE & 20 & 0.15 & 0.30 & 0.33 & 0.07 & 0.41 \\
402602 & SW & 26 & 0.35 & 0.46 & 0.44 & 0.08 & 0.41 \\
107401 & EN & 52 & 0.10 & 0.44 & 0.20 & 0.04 & 0.46 \\
409701 & SW & 20 & 0.45 & 0.45 & 0.44 & 0.09 & 0.72 \\
300649 & NL & 20 & 0.25 & 0.40 & 0.40 & $-0.06$ & 0.89 \\
105201 & EN & 37 & 0.22 & 0.38 & 0.12 & 0.08 & 0.57 \\
106901 & EN & 31 & 0.13 & 0.48 & 0.50 & $-0.01$ & 0.51 \\
105801 & EN & 24 & 0.25 & 0.46 & 0.33 & $-0.01$ & 0.56 \\
300485 & NL & 22 & 0.18 & 0.46 & 0.25 & 0.01 & 0.36 \\
407702 & SW & 20 & 0.35 & 0.40 & 0.43 & 0.07 & 0.45 \\
203202 & GE & 27 & 0.33 & 0.41 & 0.33 & 0.02 & 0.33 \\
\bottomrule
\end{tabular}
\caption{\textbf{Classes used to test the results at $\alpha=0$.} The 19 classes with $|\alpha|\leq0.1$, among the 69 of section~\ref{S-si:inference}, that lie in the region where the ESDS exists: survey unit, country, number of students~$N$, composition $(\fA,\fB,k)$, and preferences estimated with the MAM. Class 409701 is also used in the balanced-school test (Table~\ref{S-tab:C3nets}).}
\label{S-tab:esdsnets}
\end{table}

\begin{figure}[tbp]
\centering
\includegraphics[width=0.55\textwidth]{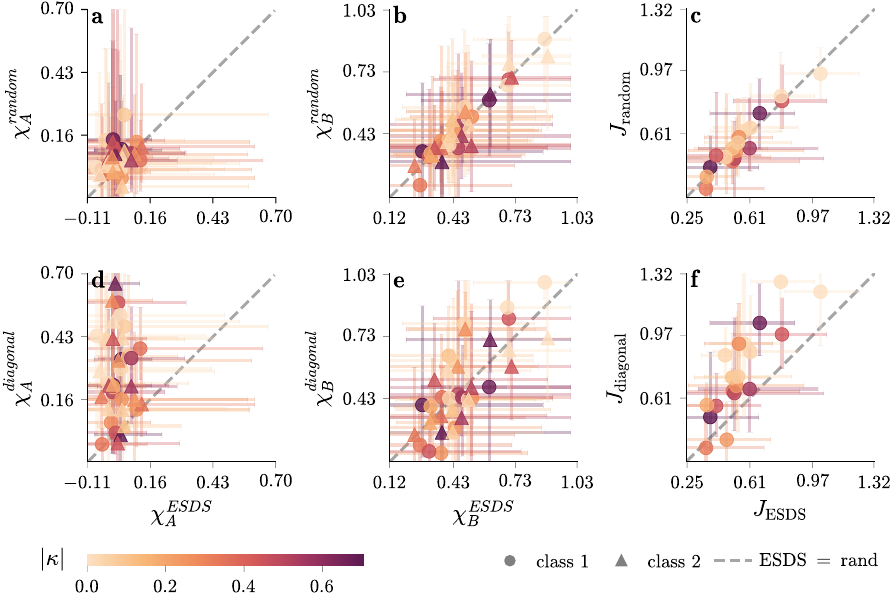}
\caption{\textbf{The ESDS, a random split and a diagonal split on the classes with $\alpha\approx0$.} Top row: (a)~class indices~$\chiA^{c_j}$ and (b)~$\chiB^{c_j}$, and (c)~aggregate objective~$J$, after the ESDS (horizontal axis) and after a random split (vertical axis); points near the diagonal indicate comparable results. Bottom row, (d)--(f): the same for the ESDS and the diagonal split, which gives a larger~$J$. Circles: class~1; triangles: class~2; color:~$|\kappa|$. Classes as in Table~\ref{S-tab:esdsnets}.}
\label{S-fig:esds_vs_random_data}
\end{figure}

\paragraph{Comparison with other splits.} We compare the ESDS with two other splits into subclasses of equal size, obtained with the same procedure and differing only in the targets $C_1[a,b]$. In the \emph{random} split, every student is assigned to either subclass at random, so both subclasses have, in expectation, the composition of the class. In the \emph{diagonal} split, subclass~1 is filled first with students of the two matching types $(A_0,B_0)$ and $(A_1,B_1)$, and then with the others, so that the two subclasses still differ in size by at most one; this pushes both subclasses away from independence. Each split is repeated 20 times. The diagonal split reinforces the correlation between the attributes and gives a larger~$J$ than the ESDS in all the classes. The comparison with the random split depends on~$|\kappa|$, as on simulated networks (section~\ref{S-si:esds_numerics}). In these classes $|\kappa|$ is moderate, since most have $k$ close to~$\fB$, and the random split and the ESDS give nearly the same results. For classes whose composition is close to independence, a random split, which requires no knowledge of the composition, therefore performs about as well as the ESDS (Fig.~\ref{S-fig:esds_vs_random_data}).

\begin{figure}[tbp]
\centering
\includegraphics[width=0.55\textwidth]{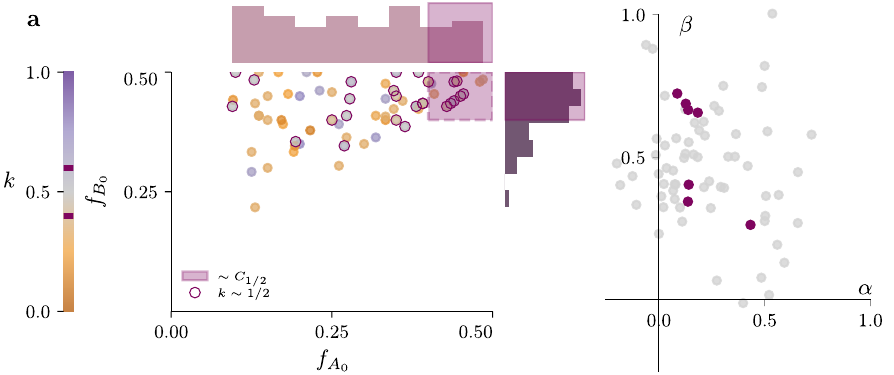}
\caption{\textbf{Classes close to the balanced school and to independence.} (a)~The CILS4EU classes in the $(\fA,\fB)$ plane, colored by~$k$, with the marginal distributions; shaded: classes close to the balanced school ($0.4\leq\fA,\fB\leq0.5$); circled: classes with $k$ close to~$1/2$. (b)~Estimated preferences $\alpha$ and~$\beta$ of all the classes (gray) and of the 7 selected classes (purple), all in the agreement phase.}
\label{S-fig:C12_selection}
\end{figure}

\paragraph{Two active preferences in a balanced school.} We select the 7 classes, among the 69, that are close to the balanced school and to independence, $0.4\leq\fA,\fB\leq0.5$ and $|k-1/2|\leq0.1$ (Fig.~\ref{S-fig:C12_selection}a and Table~\ref{S-tab:C3nets}). All of them are in the agreement phase (Fig.~\ref{S-fig:C12_selection}b), so by section~\ref{S-si:C12_transition} the best equal-size split that keeps the shares is no split, with $J=|\alpha|+|\beta|$. To test this, we give the two subclasses consolidations $k_\pm=k\pm\varepsilon$, sweep the squared offset $4\varepsilon^2$ over a uniform grid on $[0,1]$ for as long as $k_\pm$ remain within $[0,\min(1,\fB/\fA)]$, build the two subclass population matrices (checking that their average is~$F$), split the network as above, and average~$J$ over random assignments. The range of accessible offsets therefore varies across classes. The results are in Fig.~\ref{fig:C12}C of the main text.

\begin{table}[tbp]
\centering
\footnotesize
\begin{tabular}{lccccccc}
\toprule
Unit & Country & $N$ & $\fA$ & $\fB$ & $k$ & $\alpha$ & $\beta$ \\
\midrule
409701 & SW & 20 & 0.45 & 0.45 & 0.44 & 0.09 & 0.72 \\
408601 & SW & 23 & 0.43 & 0.43 & 0.60 & 0.18 & 0.66 \\
405401 & SW & 25 & 0.44 & 0.44 & 0.55 & 0.14 & 0.40 \\
103301 & EN & 25 & 0.44 & 0.48 & 0.55 & 0.43 & 0.26 \\
101702 & EN & 27 & 0.44 & 0.48 & 0.42 & 0.14 & 0.34 \\
205001 & GE & 21 & 0.43 & 0.43 & 0.44 & 0.13 & 0.69 \\
204905 & GE & 22 & 0.46 & 0.46 & 0.50 & 0.14 & 0.67 \\
\bottomrule
\end{tabular}
\caption{\textbf{Classes used to test the results in a balanced school.} The 7 classes with $0.4\leq\fA,\fB\leq0.5$ and $|k-1/2|\leq0.1$ among the 69 of section~\ref{S-si:inference}: survey unit, country, number of students~$N$, composition $(\fA,\fB,k)$, and preferences estimated with the MAM. Class 409701 is also used in the test at $\alpha\approx0$ (Table~\ref{S-tab:esdsnets}).}
\label{S-tab:C3nets}
\end{table}

\end{document}